\documentclass[useAMS,usenatbib]{mn2e}
\usepackage{aas_macros,graphicx,times,multirow,amsmath,bbold}
\usepackage[]{color}
\usepackage[draft]{hyperref}

\title[Spectrum and polarization of magnetar flares]{On the spectrum and polarization of magnetar flare emission}
\author[R. Taverna and R. Turolla]{R. Taverna,$^{1}$\thanks{E-mail:
\href{mailto:taverna@pd.infn.it}{taverna@pd.infn.it}}\smallskip \ R. Turolla$^{1,2}$\\
$^1$Department of Physics and Astronomy, University of Padova, via Marzolo 8, I-35131 Padova, Italy\\
$^2$Mullard Space Science Laboratory, University College London, Holmbury St. Mary, Surrey, RH5 6NT, UK
}

\date{Accepted \ldots. Received \ldots; in
original form \ldots} \pagerange{\pageref{firstpage}--\pageref{lastpage}} \pubyear{2017}

\def\LaTeX{L\kern-.36em\raise.3ex\hbox{a}\kern-.15em
    T\kern-.1667em\lower.7ex\hbox{E}\kern-.125emX}

\def\maxrm {\mathrm{max}}
\def\minrm {\mathrm{min}}
\def\der {\mathrm{d}}
\def\Bk {\mathrm{Bk}}
\def\echarge {\mathrm{e}}
\def\bdip {\boldsymbol{b}_\mathrm{dip}}
\def\varepsilonB {\varepsilon_\mathrm{B}}
\def\sigmaT {\sigma_\mathrm{T}}
\def\ord {\mathrm{O}}
\def\extr {\mathrm{X}}
\def\tauR {\tau_\mathrm{R}}

\begin{document}

\label{firstpage}
\maketitle
\begin{abstract}
Bursts and flares are among the distinctive observational manifestations 
of magnetars, isolated neutron stars endowed with an ultra-strong magnetic 
field ($B\approx 10^{14}$--$10^{15}$ G). It is believed that these events 
arise in a hot electron-positron plasma which remains trapped within the 
closed magnetic field lines. We developed a simple radiative transfer model 
to simulate magnetar flare emission in the case of a steady trapped fireball. 
After dividing the fireball surface in a number of plane-parallel slabs, 
the local spectral and polarization properties are obtained integrating 
the radiative transfer equations for the two normal modes. We assume that 
magnetic Thomson scattering is the dominant source of opacity, and neglect 
contributions from second-order radiative processes, although double-Compton 
scattering is accounted for in establishing local thermal equilibrium in 
the fireball atmospheric layers. The observed spectral and polarization 
properties as measured by a distant observer are obtained summing the contributions 
from the patches which are visible for a given viewing geometry by means 
of a ray-tracing code. The spectra we obtained in the $1$--$100$ keV energy 
range are thermal and can be described in terms of the superposition of 
two blackbodies. The blackbody temperature and the emitting area ratio are 
in broad agreement with the available observations. The predicted linear 
polarization degree is in general greater than $80\%$ over the entire energy 
range and should be easily detectable by 
new-generation X-ray polarimeters, like IXPE, XIPE and eXTP.
\end{abstract}

\section{Introduction}
\label{intro}
Soft gamma repeaters (SGRs) and anomalous X-ray pulsars (AXPs) are regarded
as the observational manifestations of the same class of neutron stars (NSs), 
aka the magnetars, for which the long measured spin periods ($P\approx 2$--$12$ 
s) and large period derivatives ($\dot{P}\approx 10^{-13}$--$10^{-10}$ ss$^{-1}$) 
lead to a huge value of the dipole magnetic field, $B\approx 10^{14}$--$10^{15}$ 
G, well above those of other NS classes \cite[see][for reviews]{tzw15,mer08}. 
Magnetars show an X-ray persistent emission with luminosity in the range 
$10^{33}$--$10^{36}$ ergs$^{-1}$, some orders of magnitude greater than 
the spin-down luminosity $\dot{E}_\mathrm{rot}$ inferred from $P$ and $\dot{P}$. 
One of the most distinctive properties of SGRs/AXPs is the emission of short 
($\approx 10^{-2}$--$1$ s), energetic ($\approx 10^{36}$--$10^{41}$ erg) 
X-ray bursts and longer ($\approx 1$--$50$ s), even more energetic ($\approx 
10^{41}$--$10^{43}$ erg) intermediate flares. Furthermore, three SGRs have 
been observed to emit also giant flares, the most powerful events ever observed 
from compact objects, characterized by a short ($\approx0.1$--$1$ s) initial 
spike, followed by a long ($\approx10^2$--$10^3$ s) pulsating tail modulated 
at the spin frequency of the star, with a total energy release $\approx 
10^{44}$--$10^{47}$ erg.

According to the magnetar model, firstly developed by \citet{dt92}, magnetar
activity is sustained by the magnetic energy stored in the huge (internal) 
magnetic field. The latter is believed to develop a large toroidal component 
\cite[see e.g.][]{brait09,pp11}, able to exert a strong magnetic stress 
on the conductive star crust. Contrary to what happens in ``normal'' NSs, 
where this force can be balanced by the rigidity of the crust, in the case 
of magnetars the internal stresses are strong enough to displace single 
surface elements (the so-called starquakes), owing to their ultra-strong 
fields. As a result, the external magnetic field acquires in turn a non-zero 
toroidal component, becoming twisted, and this makes possible for charged 
particles to fill the magnetosphere, streaming along the closed field lines 
\cite[][]{tlk02,ntz08,tzw15}. 

The mechanisms that trigger magnetar bursting activity are still not completely 
clear. It has been proposed \cite[see][see also \citealt{elenbass16} for a discussion]{lyut03,woodsetal05} that fast 
acceleration of magnetospheric particles, after spontaneous magnetic field 
reconnections, could be responsible for the enormous energy release and 
the light curves observed in particular kinds of SGR bursts. \citet{td95,td01} 
suggested an alternative model to explain the giant flare trigger and emission 
mechanisms. According to them, an internal magnetic field instability induces 
large-amplitude oscillations in the magnetosphere, that convert, in turn, 
into a hot electron-positron plasma. While part of this plasma quickly escapes 
outwards in the initial phases, producing the hard initial peak of giant 
flares, another part remains trapped within the closed magnetic field lines, 
resulting in an optically-thick, photon-pair fireball. The long pulsating 
tail observed in these extreme events should be due indeed to the radiation 
coming from this confined, cooling region in the magnetosphere. The same 
paradigm can be also used to properly explain the shorter intermediate flares, 
which show as well a clearly detectable decay tail \cite[][]{olive04,feretal04,isretal08}.

Although many efforts have been made in the last decades to explain the 
physics of magnetars, a detailed theoretical framework for modeling the 
spectral and polarization properties of the radiation emitted during magnetar 
flares does not yet exist. Spectral properties are difficult to investigate 
due to the rarity of intermediate/giant flares and the extremely small duration 
times of short bursts. On the other hand, as shown e.g. by \citet{isretal08}, 
the occurrence of batches of these events, during which a great number of 
single short bursts and intermediate flares can be emitted, could compensate 
for their short duration and provide enough statistics for a meaningful 
analysis. In the case of mangetars also polarimetry, besides spectral analysis, 
can be profitably used to characterize the observed radiation and identify 
its production mechanisms. In fact, in the presence of strong magnetic fields 
photons are expected to be highly polarized in two normal modes, the ordinary 
and the extraordinary modes. The radiative processes that take place in 
the magnetosphere generally influence the polarization pattern of the emitted 
radiation \cite[see][]{mesz92,ntz08}. Moreover, for magnetic field intensities 
high enough, the photon polarization is modified also when they propagate 
in vacuo \cite[the vacuum birefringence effect, see][see also \citealt{heylsh02,mignetal17} 
and references therein]{heiseul36}. This kind of analysis has not been possible 
so far, given that no instruments are available to perform polarization 
measurements in the X-rays. Nevertheless, new impetus was given in this 
field by new-generation instruments like IXPE \cite[][recently approved 
for the NASA SMEX program and to be launched within 2020]{weiss13}, XIPE 
\cite[][in the study phase of ESA M4 program]{soff16} and eXTP \cite[][]{zhang16}, 
which promise to open a new window in X-ray astrophysics.

The problem of radiative transfer in a scattering medium in the presence 
of strong magnetic fields, as well as its possible applications to the spectra 
of SGR bursts, have been addressed by \citet{lyub02}. Starting from the 
same theoretical framework as in \citet{td95,td01}, he calculated the spectra 
of radiation escaping from a plane-parallel slab of the pair fireball, in 
one dimensional approximation and assuming the star magnetic field parallel 
to the slab normal. More recently, \citet{yz15} have presented Montecarlo 
simulations to study the polarization properties of giant flare decay tail 
emission. They asssumed that the pair plasma produced during the magnetar 
flare remains trapped within a set of closed (dipolar) magnetic field lines 
and solved the radiative transport in a geometrically-thin, surface layer 
of the fireball, where magnetic Thomson scattering is the only source of 
opacity. According to the results of their simulations, radiation emitted 
during such events would be only mildly polarized, with maximum polarization 
degree in between $\approx10$ and $30\%$ depending on the photon energy. 
The spectrum and polarization of magnetar flare radiation have been also 
studied by \citet{vput16}. While considering as well the effects of magnetic 
scattering on photons that propagate through a trapped fireball (modelled 
in a similar way as in \citealt{yz15}), they focused in particular on the 
radiation beaming driven by the presence of a mildly-relativistic, baryon-loaded 
outflow outside the fireball \cite[predicted in the model by][]{td95,td01}, 
in order to describe consistently the observed pulse profiles and the expected 
evolution of a magnetar fireball. However, a detailed analysis of the spectral 
and polarization properties of the emitted radiation was outside the scope 
of their work.

In this paper we reconsider the problem of both the spectrum and polarization
of the radiation emitted from a steady trapped fireball, providing simulations 
directly comparable with observations. As in \citet{yz15}, we compute the 
photon transport in the surface layer of the fireball, following the approach 
by \citet{lyub02} and assuming magnetic Thomson scattering as the dominant 
source of opacity in the plasma. We neglect the contributions from second-order 
radiative processes, although the presence of double-Compton scattering 
is accounted for in establishing local thermal equilibrium in the fireball 
atmospheric layers. In particular, after dividing the fireball surface in 
a number of plane-parallel slabs, we obtain the local spectral and polarization 
properties integrating the radiative transfer equations for the two normal 
modes. The observed spectral and polarization properties as measured by 
a distant observer are obtained summing the contributions from the patches 
which are visible for a given viewing geometry by means of a ray-tracing 
code. Our results show that the simulated spectra in the $1$--$100$ keV 
energy range can be suitably described in terms of the superposition of 
two blackbody components, with temperatures and emitting area ratio in broad 
agreement with the observations available so far \cite[see e.g.][]{isretal08,olive04,feretal04}. 
Furthermore, the predicted polarization pattern significantly differs from 
those presented in previous works. In fact, the linear polarization degree
turns out to be in general greater than $80\%$ over the entire energy range. 
Such a large degree of polarization should be easily detectable by new-generation 
X-ray polarimeters, allowing to confirm the model predictions. We stress, 
however, that our model is focussed on the radiation coming from a steady
trapped fireball, and a complete study of how magnetar flares rise and fade 
away is beyond the scope of this paper. 

The outline of the paper is as follows. In section \ref{section:theoreticalmodel}
we introduce the theoretical framework of our model, discussing the scattering
cross sections and the radiative transfer equations. In section \ref{section:numericalimplementation}
we describe the structure of the radiation transfer and the ray-tracing 
codes we developed and discuss the visibility of a trapped fireball in the 
magnetosphere of a magnetar. The results of our simulations are presented 
in section \ref{section:simulationsresults}, while discussion and conclusions 
are reported in section \ref{section:discussion}.

\section{Theoretical model}
\label{section:theoreticalmodel}

According to the model originally developed by \citet{dt92,td95}, magnetar
flares originate in the sudden rearrangements of the external magnetic field, 
triggered by crustal displacements driven by their strong internal field; 
such events are able to inject an Alfv\'{e}n pulse into the magnetosphere. 
If the magnetic field at a certain distance $R_\maxrm$ from the star is 
still strong enough to contain the energy of the wave in a volume $\sim
R_\maxrm^3$, these Alfv\'{e}n waves remain trapped within the closed field 
lines characterized by the maximum radius $R_\maxrm$, dissipating into a 
magnetically confined electron-positron plasma and forming a so-called ``trapped 
fireball'' \cite[see e.g.][and references therein]{td95,td01}. Here we focus 
on a simple model to compute the properties of radiation emitted during 
a typical magnetar flare, by solving the radiative transfer equation in 
a pure scattering medium, following the approach by \citet{lyub02}. In some 
respects, our model is similar to that discussed by \citet{yz15}, inasmuch 
we consider only a steady trapped fireball, where the scattering optical 
depth is expected to be very high; this allows to compute photon transport 
only in a geometrically-thin surface slab. Our results can be then used 
as inputs in more sophisticated models which e.g. account for particle outflows.

\subsection{Scattering in strong magnetic fields}
\label{subsection:crosssections}
In the presence of magnetar-like magnetic fields, photons are expected to 
be linearly polarized in two normal modes \cite[e.g.][]{gnpa74,holai03,laietal10}: 
the ordinary mode (O), with the polarization vector lying in the plane defined 
by the photon propagation direction $\boldsymbol{k}$ and the local magnetic 
field $\boldsymbol{B}$, and the extraordinary mode (X), with the electric 
field oscillating perpendicularly to both $\boldsymbol{k}$ and $\boldsymbol{B}$. 
Moreover, the electron cyclotron energy $\varepsilonB=m_\mathrm{e}c^2B/B_\mathrm{Q}$ 
(here $m_\mathrm{e}$ is the electron mass, $B_\mathrm{Q}=m_\mathrm{e}^2c^3/\hbar\echarge 
\simeq 4.414\times 10^{13}$ G the quantum critical field and $\echarge$  
the electron charge) is typically above the energy of the photons emitted
during a magnetar burst, so that scattering onto electrons/positrons is 
non-resonant. In the electron rest frame (ERF) and neglecting the charge 
recoil, the scattering cross sections depend on the polarization state of 
both the ingoing and the outgoing photons \cite[see e.g.][]{her79,vent79,mesz92},
\begin{flalign} \label{equation:crosssecBnpz}
\sigma_\mathrm{OO}(\alpha\rightarrow\alpha') &= \dfrac{3}{8\pi}(1-\mu^2_\Bk)(1-\mu'^2_\Bk)\delta(\varepsilon'-\varepsilon) \nonumber & \\
\sigma_\mathrm{OX}(\alpha\rightarrow\alpha') &= \dfrac{3}{8\pi}\left(\dfrac{\varepsilon}{\varepsilonB}\right)^2\mu^2_\Bk\cos^2(\phi_\Bk-\phi'_\Bk)\delta(\varepsilon'-\varepsilon) \nonumber & \\
\sigma_\mathrm{XO}(\alpha\rightarrow\alpha') &= \dfrac{3}{8\pi}\left(\dfrac{\varepsilon}{\varepsilonB}\right)^2\mu'^2_\Bk\cos^2(\phi_\Bk-\phi'_\Bk)\delta(\varepsilon'-\varepsilon) \nonumber & \\
\sigma_\mathrm{XX}(\alpha\rightarrow\alpha') &= \dfrac{3}{8\pi}\left(\dfrac{\varepsilon}{\varepsilonB}\right)^2\sin^2(\phi_\Bk-\phi'_\Bk)\delta(\varepsilon'-\varepsilon)\,, &
\end{flalign}
where a prime labels the quantities after scattering, $\varepsilon$ is the 
photon energy, $\mu_\Bk$ is the cosine of the angle $\theta_\Bk$ between 
the photon direction and the local magnetic field and $\phi_\Bk$ is the 
associated azimuth. Here we introduced the notation
\begin{flalign} \label{equation:sigmafreccia}
\sigma_{ij}(\alpha\rightarrow\alpha') & \equiv \frac{1}{\sigmaT}\left[\frac{\der^2\sigma}{\der\varepsilon'\der\Omega'}\right]_{ij}\,, &
\end{flalign}
where $i,j=\mathrm{O},\mathrm{X}$, $\der\Omega'=\der\mu'_\Bk\der\phi'_\Bk$ 
and $\sigmaT$ is the Thomson cross section. The previous expressions hold 
as far as the vacuum contributions in the dielectric tensor dominate over 
the plasma ones \cite[see e.g.][and references therein]{hl06}. Equations 
(\ref{equation:crosssecBnpz}) show that photons can change their initial 
polarization state upon scattering. Furthermore, it appears clearly that 
all the cross sections which involve extraordinary photons are suppressed 
by a factor $(\varepsilon/\varepsilonB)^2\propto(\varepsilon\,B_\mathrm{Q}/B)^2$ 
with respect to the O-O cross section, that is essentially of the order 
of $\sigmaT$. This implies that the medium becomes optically thin for X-mode 
photons at much larger Thomson depths with respect to O-mode photons.

\begin{figure*}
\begin{center}
\includegraphics[width=15cm]{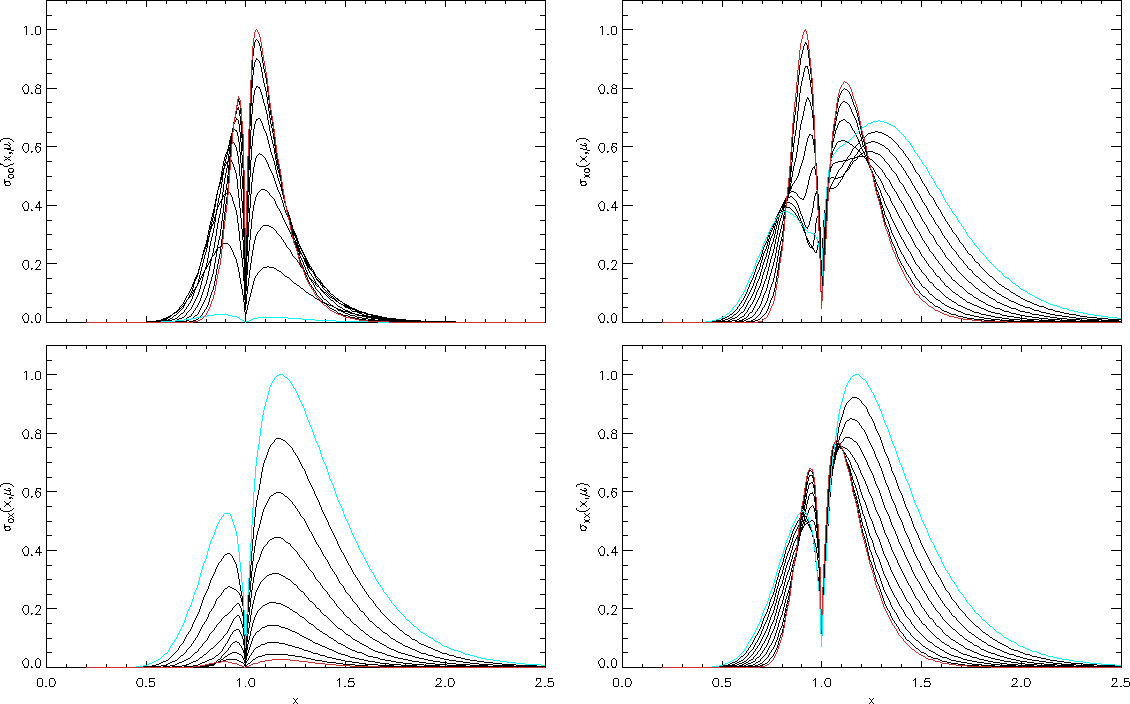}
\caption{Normalized cross sections for non-resonant, magnetic Compton scattering, 
integrated over the solid angle $\Omega'$ of the outcoming photons, plotted 
as functions of the ratio $x=\epsilon'/\epsilon$ of the ingoing and outgoing 
photon energies. The curves refer to different values of $\cos\vartheta_\Bk$
ranging from 1 (red) to 0.1 (light-blue) with step 0.1 (the integrated cross 
sections turn out to be symmetric replacing $\theta_\Bk$ with $\pi-\theta_\Bk$).}
\label{figure:integratedsigma}
\end{center}
\end{figure*}
Equations (\ref{equation:crosssecBnpz}) describe correctly the scattering
process in a medium where electrons (and positrons) are substantially at 
rest in the stellar frame. For the sake of completeness, we discuss here 
also the general case of electrons and positrons moving at speed $\beta$ 
(in unit of the speed of light $c$), still treating the scattering in the 
particle rest frame as conservative. In this case, the expressions for the 
scattering cross sections in the stellar frame can be obtained from those 
in the particle rest frame, given by equations (\ref{equation:crosssecBnpz}), 
through the transformation \cite[see][]{pomr73}
\begin{flalign} \label{equation:initialNCcrosssec}
\bar{\sigma}_{ij}&=\frac{D}{D'}\sigma_{ij}\,, &
\end{flalign}
where
\begin{flalign} \label{equation:DDprime}
&D=1-\beta\cos\vartheta_\Bk\,\,\,\,\,\,\,\,\,\,\,\,D'=1-\beta\cos\vartheta'_\Bk &
\end{flalign}
and $\vartheta_\Bk$ ($\vartheta'_\Bk$) is the angle between the incoming
(outcoming) photon direction and the stellar magnetic field in the star
frame. In particular, the angles in the stellar frame can be related to 
those in the particle frame using the angular aberration formula,
\begin{flalign} \label{equation:aberrationformula}
&\mu_\Bk=\frac{\cos\vartheta_\Bk-\beta}{D}\,\,\,\,\,\,\,\,\,\,\,\,
\mu'_\Bk=\frac{\cos\vartheta'_\Bk-\beta}{D'}&
\end{flalign}
while, for the energy, it is
\begin{flalign} \label{equation:energyboost}
&\varepsilon=\gamma D\epsilon\,\,\,\,\,\,\,\,\,\,\,\,\varepsilon'=\gamma D'\epsilon'\,, &
\end{flalign}
with $\epsilon$ ($\epsilon'$) the incoming (outcoming) photon energy in 
the star reference frame and $\gamma$ the particle Lorentz factor.

Integrating equation (\ref{equation:initialNCcrosssec}) in the velocity
space of the scattering particles gives the Compton scattering kernel in 
the stellar frame,
\begin{flalign} \label{equation:NCCSK}
\bar{\sigma}_{ij}(\epsilon\rightarrow\epsilon',\Omega\rightarrow\Omega')&=
\int\sigma_{ij}\frac{n_\mathrm{e}}{\gamma}\frac{D}{D'}f(\beta)\der\beta\,, &
\end{flalign}
in the case of an isotropic velcity distribution. Here $n_\mathrm{e}$ is the particle number density and $f(\beta)$ is the 
particle velocity distribution, which we assume to be a relativistic (1D) 
maxwellian
\begin{flalign} \label{equation:MaxwellJuttner}
f(\beta)\der\beta&=\frac{\gamma^3\mathrm{e}^{-(\gamma-1)/\bar{\Theta}}\der\beta}{2\mathrm{e}^{1/\bar{\Theta}}K_1(1/\bar{\Theta})}\,, &
\end{flalign}
with $\bar{\Theta}\equiv kT/m_\mathrm{e}c^2$ ($T$ is the plasma temperature)
and $K_1$ the modified Bessel function of the second kind. Finally, using 
equations (\ref{equation:crosssecBnpz}) and (\ref{equation:DDprime}) -- 
(\ref{equation:energyboost}), together with the properties of the $\delta$-function, 
equation (\ref{equation:NCCSK}) becomes
\begin{flalign} \label{equation:NCcrosssec}
\bar{\sigma}_{ij}(\alpha\rightarrow\alpha') &= \frac{n_\mathrm{e}|\beta|e^{-(\gamma-1)/\bar{\Theta}}}{2\mathrm{e}^{1/\bar{\Theta}}K_1(1/\bar{\Theta})}\Sigma_{ij}\,, &
\end{flalign}
where
\begin{flalign} \label{equation:NCSigmabig}
\Sigma_\mathrm{OO} &\equiv \frac{x(1-\cos^2\vartheta_\Bk)}{\epsilon|x-1|}\frac{(1-\cos^2\vartheta'_\Bk)}{\gamma^3(1-\beta\cos\vartheta_\Bk)^2(1-\beta\cos\vartheta'_\Bk)^2} & \nonumber \\
\Sigma_\mathrm{OX} &\equiv \frac{x(\cos\vartheta_\Bk-\beta)^2}{\epsilon|x-1|}\left(\frac{\epsilon B_\mathrm{Q}}{B}\right)^2\gamma^3\cos^2(\phi_\Bk-\phi'_\Bk) \nonumber & \\
\Sigma_\mathrm{XO} &\equiv \frac{x^3(\cos\vartheta'_\Bk-\beta)^2}{\epsilon|x-1|}\left(\frac{\epsilon B_\mathrm{Q}}{B}\right)^2\gamma^3\cos^2(\phi_\Bk-\phi'_\Bk) \nonumber & \\
\Sigma_\mathrm{XX} &\equiv \frac{x(1-\beta\cos\vartheta_\Bk)^2}{\epsilon|x-1|}\left(\frac{\epsilon B_\mathrm{Q}}{B}\right)^2\gamma^3\sin^2(\phi_\Bk-\phi'_\Bk) \nonumber & \\
\end{flalign}
and $x\equiv\epsilon'/\epsilon$. As an example, Figure \ref{figure:integratedsigma} 
shows the behavior of the (normalized) cross sections (\ref{equation:NCcrosssec}), 
integrated over the outgoing photon direction, as a function of $x$, for 
different values of $\cos\vartheta_\Bk$ and for $\phi_\Bk=0$.

\subsection{Radiative transfer in the fireball atmosphere}
\label{subsection:radiativetransfer}
In this work we treat the fireball plasma as a pure-scattering medium. Since 
our primary goal in this investigation is to provide and test a general 
method to compute burst spectral and polarization properties, in the following 
we restrict to Thomson scattering. The full case of Compton scattering will 
be addressed in a sequel paper. Among the additional second-order processes 
that can take place in magnetized plasma, we focus on the effects of double-Compton 
scattering only, since the contributions of other processes, such as photon 
splitting and thermal bremsstrahlung, turn out to be far less important 
under the assumptions we made (see section \ref{section:discussion} for 
a more complete discussion).

Double-Compton scattering is, then, the main process responsible for photon 
production in the fireball medium. It has been shown \cite[see][and references 
therein]{lyub02} that, at photon energies $\varepsilon\ll\varepsilonB$, 
this process can be treated as non-resonant, resembling its non-magnetic 
counterpart, even when a magnetic field is present. Low-energy photons ($\varepsilon\ll 
kT$) are injected in the fireball due to double-Compton scattering, at a 
rate \cite[see][]{light81}
\begin{flalign} \label{equation:Qdouble}
Q &\approx \frac{4\alpha_\mathrm{F}}{3\pi}\frac{\sigmaT}{m_e^2c^4}\frac{\exp(\varepsilon/kT)-1}{\varepsilon^3}\left[f_\mathrm{B}(\varepsilon,T)-f(\varepsilon)\right]I\,, &
\end{flalign}
where $\alpha_\mathrm{F}$ is the fine-structure constant, $f(\varepsilon)$ 
is the photon occupation number, $f_\mathrm{B}(\varepsilon,T)=[\exp(\varepsilon/kT)-1]^{-1}$ 
and
\begin{flalign}
I &\equiv \int\varepsilon^4[1+f(\varepsilon)]f(\varepsilon)\der\varepsilon\,. &
\end{flalign}
For large scattering depths, equation (\ref{equation:Qdouble}) ensures that 
photons follow a Planck distribution at photon energies low enough to make 
double-Compton scattering dominant, $\varepsilon\ll\varepsilon_0\approx 0.03(kT)^{3/2}$. 
At higher energies, on the other hand, scattering tends to establish a Bose-Einstein 
distribution $f_\mathrm{BE}(\varepsilon,T)=[\exp(\varepsilon+\mu/kT)-1]^{-1}$.
However, as shown by \citet{lyub02}, the chemical potential $\mu$ satisfies 
the condition
\begin{flalign}
\ln\frac{\mu+\varepsilon_0}{\varepsilon_0}&\ll0.5\bigg(\frac{10B_\mathrm{Q}}{B}\bigg)^2\,, &
\end{flalign}
where the right-hand side is calculated at scattering depth unity. In this
way $\mu$ remains small and one can solve the photon transport assuming 
local thermal equilibrium (LTE) at large depths for both O- and X-mode photons.

We solved the radiative transfer equations for both the ordinary and the 
extraordinary photons in the geometrically-thin, surface layers of the fireball 
which we term the atmosphere. The latter is divided into a number of patches, 
each labelled by the intensity of the magnetic field at the patch centre 
and by the angle $\theta_\mathrm{B}$ that $\boldsymbol{B}$ makes with the 
local normal $\boldsymbol{z}$. The contributions from each patch in view 
are then summed together to derive the overall spectral and polarization 
properties of the emitted radiation. We assume that the patch dimensions 
are small enough with respect to the radial scale ($\approx$ the star radius) 
to neglect the curvature of the closed field lines that contain the fireball. 
This approach allows us to treat the atmosphere in the plane-parallel approximation, 
i.e. all the relevant quantities depend only on the height $z$ with respect 
to the base of the layer. At variance with the case discussed by \citet{lyub02}, 
in our model the magnetic field is not necessarily aligned with the patch 
normal. When this happens, the angle $\theta_z$ between the photon direction 
and $\boldsymbol{z}$ differs from $\theta_\mathrm{Bk}$ (see section \ref{subsection:crosssections}),
and the two angles are related by
\begin{flalign} \label{equation:muz}
\mu_z &= \mu_\Bk\mu_\mathrm{B}-\sqrt{(1-\mu^2_\Bk)(1-\mu^2_\mathrm{B})}\cos\phi_\mathrm{Bk}\,, &
\end{flalign}
with $\mu_z=\cos\theta_z$ and $\mu_\mathrm{B}=\cos\theta_\mathrm{B}$.

In a pure scattering medium, the radiative transfer equations written in 
terms of the photon number intensity $n_i$ ($i=\mathrm{O,X}$) take the form 
\cite[see e.g.][]{meszetal89,alexetal89},
\begin{flalign} \label{equation:initRTE}
\mu_z\frac{\der n_i}{\der\tau}&= \sum_{k=\mathrm{O,X}}\int\bigg\{-\sigma_{ik}(\alpha\rightarrow\alpha')n_i(\alpha)\big[1+n_k(\alpha')\big] & \nonumber \\
\ &\,\,\,\,\,\,+\sigma_{ki}(\alpha'\rightarrow\alpha)\bigg(\frac{\varepsilon'}{\varepsilon}\bigg)^2n_k(\alpha')\big[1+n_i(\alpha)\big]\bigg\}\der\varepsilon'\der\Omega'\,, &
\end{flalign}
where stimulated scattering is accounted for. Here \mbox{$\der\tau=n_\mathrm{e}\sigmaT\der s$} 
is the infinitesimal Thomson depth, with $s$ a parameter along the photon 
propagation direction. However the previous expression can be considerably 
simplified. In fact, in LTE the scattering cross sections 
(\ref{equation:crosssecBnpz}) must obey the detailed balance condition \cite[see 
e.g.][]{meszetal89,alexetal89,alexmesz91},
\begin{flalign} \label{equation:detailedbalance}
\sigma_{ki}(\alpha'\rightarrow\alpha) &=\bigg(\frac{\varepsilon}{\varepsilon'}\bigg)^2\exp\big[-(\varepsilon-\varepsilon')/kT\big]\sigma_{ik}(\alpha\rightarrow\alpha')\,; &
\end{flalign} 
in this way, equation (\ref{equation:initRTE}) becomes
\begin{flalign} \label{equation:medRTE}
\mu_z\frac{\der n_i}{\der\tau} &= \sum_{k=\mathrm{O,X}}\int\bigg\{-\sigma_{ik}(\alpha\rightarrow\alpha')n_i(\alpha) & \nonumber \\
\ &\,\,\,\,\,\,+\sigma_{ik}(\alpha\rightarrow\alpha')F_i(\alpha,\varepsilon')n_k(\alpha')\bigg\}\der\varepsilon'\der\Omega'\,, &
\end{flalign}
where
\begin{flalign} \label{equation:capitalF}
F_i(\alpha,\varepsilon')&\equiv\exp\big[-(\varepsilon-\varepsilon')/kT\big]\bigg(1+n_i(\alpha)\bigg)-n_i(\alpha)\,. &
\end{flalign}
Finally, under the assumption of conservative scattering ($\varepsilon=\varepsilon'$)
and substituting the expressions (\ref{equation:crosssecBnpz}), one obtains
\begin{flalign} \label{equation:finRTE}
\mu_z\frac{\der n_\ord}{\der\tau} &= -\bigg[1-\mu^2_\Bk+\frac{3\mu^2_\Bk}{4}\bigg(\frac{\varepsilon}{\varepsilonB}\bigg)^2\bigg]n_\ord(\alpha) & \nonumber \\
\ &\,\,\,\,\,\,+\frac{3}{8\pi}\int_{4\pi}\bigg[(1-\mu_\Bk^2)(1-\mu_\Bk'^2)n_\ord(\alpha') & \nonumber \\
\ &\,\,\,\,\,\,+\bigg(\frac{\varepsilon}{\varepsilonB}\bigg)^2\mu^2_\Bk\cos^2(\phi_\Bk-\phi'_\Bk)n_\extr(\alpha')\bigg]\der\Omega' \nonumber & \\
\ &\ \nonumber & \\
\mu_z\frac{\der n_\extr}{\der\tau} &= -\bigg(\frac{\varepsilon}{\varepsilonB}\bigg)^2n_\extr(\alpha) & \nonumber \\
\ &\,\,\,\,\,\,+\frac{3}{8\pi}\bigg(\frac{\varepsilon}{\varepsilonB}\bigg)^2\int_{4\pi}\bigg[\sin^2(\phi_\Bk-\phi'_\Bk)n_\extr(\alpha') & \nonumber \\
\ &\,\,\,\,\,\,+\mu'^2_\Bk\cos^2(\phi_\Bk-\phi'_\Bk)n_\ord(\alpha')\bigg]\der\Omega'\,. &
\end{flalign}

As discussed in the previous section, the propagation of photons in the 
fireball medium is quite different according to their polarization mode.
These two different behaviors can be described in terms of the O- and X-mode
optical depths $\tau_i\sim n_\mathrm{e}\sigma_iH$ ($i=\mathrm{O,\, X}$), 
where $\sigma_i=\int(\sigma_{i\mathrm{O}}+\sigma_{i\mathrm{X}})\der\varepsilon'\der\Omega'$ 
and $H$ is the scale height of the fireball atmospheric layer. Taking into 
account the cross sections (\ref{equation:crosssecBnpz}) it results
\begin{flalign} \label{equation:tauOtauX}
\tau_\ord&\approx\tau & \nonumber \\  
\tau_\extr&=\left(\frac{\varepsilon}{\varepsilonB}\right)^2\tau_\ord\,, &
\end{flalign}
with $\tau$ the Thomson scattering depth. We remind that these considerations and the following equations
hold in the limit $\varepsilon\ll \varepsilonB$. Owing to the suppression factor 
$(\varepsilon/\varepsilonB)^2$ of the X-mode photon cross section with respect 
to the O-mode one (see section \ref{subsection:crosssections}), the photosphere 
of X-mode photons (i.e. the layer at which $\tau_\extr\approx 1$) lies at 
different heights in the fireball atmosphere for photons with different 
energies; in particular, it is closer to the top the higher the photon energy. 
At the same time, it is $\tau_\ord\gg 1$ at the X-mode photosphere for all
the energies of interest ($\sim$ 1--100 keV).

For these reasons, we follow the approach described by \citet{lyub02} and 
solve the photon transport in terms of the Rosseland mean optical depth 
$\tauR$ for the X-mode photons, defined by
\begin{flalign} \label{equation:tauRdef}
\der\tauR&= n_\mathrm{e}\sigma_\mathrm{X,R}\der s\,, &
\end{flalign}
where $\sigma_\mathrm{X,R}$ is the Rosseland mean of the $\sigma_\extr$ 
cross section
\begin{flalign} \label{equation:Rossmeanlambda}
\sigma_\mathrm{X,R}&=\left[\frac{\int\sigma_\extr^{-1} \partial B_\varepsilon/\partial T\,\der\varepsilon}{\int\partial B_\varepsilon/\partial T\,\der\varepsilon}\right]^{-1} & \nonumber \\
 &= \frac{4\pi^2}{5}\sigmaT\left(\frac{kTB_\mathrm{Q}}{m_\mathrm{e}c^2B}\right)^2\,, &
\end{flalign}
with $B_\varepsilon(T)=\varepsilon^3/[\exp(\varepsilon/kT)-1]$. A 
good approximation for the temperature distribution can be obtained, in 
the diffusion regime ($\tau>1$), for a scattering dominated medium \cite[see][and 
references therein]{lyub02},
\begin{flalign} \label{equation:temperaturedistrib}
T&=T_\mathrm{b}\sqrt{1+\frac{3}{4}\tauR}\,. &
\end{flalign}
where $T_\mathrm{b}$ is the bolometric temperature\footnote{$T_\mathrm{b}$ 
is defined in terms of the total radiation flux $\mathcal{F}$ as $T_\mathrm{b}=(\mathcal{F}/\sigma)^{1/4}$, 
with $\sigma$ the Stefan-Boltzmann constant.}. Actually, solving the temperature 
profile in this particular case is a non-trivial problem, since it involves 
the computation of the energy exchange between the radiation field and the 
pair plasma in the presence of double-Compton and non-conservative scattering.
Strictly speaking, the validity of equation (\ref{equation:temperaturedistrib}) 
is restricted to the optically thick limit. However, it provides a good 
approximation to the numerical solution one obtains solving the energy balance 
in the fireball medium also at small optical depths, as we checked a posteriori. 
Deviations turn out to be $\la 15\%$, in agreement with what found also 
by \citet{lyub02}. The Rosseland mean optical depth follows from equations 
(\ref{equation:Rossmeanlambda}) and (\ref{equation:temperaturedistrib}),
\begin{flalign} \label{equation:Rossmeandepth}
\tauR&=\frac{4\pi^2}{5}\left(\frac{kT_\mathrm{b}B_\mathrm{Q}}{m_\mathrm{e}c^2B}\right)^2\sigmaT\int n_\mathrm{e}\der s = R(B)\tau\,, &
\end{flalign}
where
\begin{flalign} \label{equation:RenFac}
R(B)&\equiv\frac{4\pi^2}{5}\left(\frac{kT_\mathrm{b}B_\mathrm{Q}}{m_\mathrm{e}c^2B}\right)^2\,. &
\end{flalign}
This allows to relate the optical depths $\tau_\ord$ and $\tau_\extr$ 
\begin{flalign}
\tau_\ord&=\frac{5}{4\pi^2}\left(\frac{m_\mathrm{e}c^2B}{kT_\mathrm{b}B_\mathrm{Q}}\right)^2\tauR\,,\,\,\,\,\,\,\,\,\,\,\,\,\,\,\,
\tau_\extr=\frac{5}{4\pi^2}\left(\frac{\varepsilon}{kT_\mathrm{b}}\right)^2\tauR\,. &
\end{flalign}
Finally, using equation (\ref{equation:Rossmeandepth}) one can rewrite the 
radiative transfer equations (\ref{equation:finRTE}) in terms of the Rosseland 
mean optical depth simply scaling the right-hand sides of both of them by 
the factor $R(B)$:
\begin{flalign} \label{equation:finRTEtauR}
\mu_z\frac{\der n_\ord}{\der\tauR} &= -\frac{1}{R(B)}\bigg\{\bigg[1-\mu^2_\Bk+\frac{3\mu^2_\Bk}{4}\bigg(\frac{\varepsilon}{\varepsilonB}\bigg)^2\bigg]n_\ord(\alpha) & \nonumber \\
\ &\,\,\,\,\,\,+\frac{3}{8\pi}\int_{4\pi}\bigg[(1-\mu_\Bk^2)(1-\mu_\Bk'^2)n_\ord(\alpha') & \nonumber \\
\ &\,\,\,\,\,\,+\bigg(\frac{\varepsilon}{\varepsilonB}\bigg)^2\mu^2_\Bk\cos^2(\phi_\Bk-\phi'_\Bk)n_\extr(\alpha')\bigg]\der\Omega'\bigg\} \nonumber & \\
\ &\ \nonumber & \\
\mu_z\frac{\der n_\extr}{\der\tauR} &= -\frac{1}{R(B)}\bigg\{\bigg(\frac{\varepsilon}{\varepsilonB}\bigg)^2n_\extr(\alpha) & \nonumber \\
\ &\,\,\,\,\,\,+\frac{3}{8\pi}\bigg(\frac{\varepsilon}{\varepsilonB}\bigg)^2\int_{4\pi}\bigg[\sin^2(\phi_\Bk-\phi'_\Bk)n_\extr(\alpha') & \nonumber \\
\ &\,\,\,\,\,\,+\mu'^2_\Bk\cos^2(\phi_\Bk-\phi'_\Bk)n_\ord(\alpha')\bigg]\der\Omega'\bigg\}\,. &
\end{flalign}

\section{Numerical implementation}
\label{section:numericalimplementation}
In this section we illustrate the numerical method we used to compute the
spectra of the photons emitted from a typical magnetar fireball. Hereafter 
we will refer to a template neutron star with radius $R_\mathrm{NS}=10$ 
km and mass $M_\mathrm{NS}=1.4\,M_\mathrm{\odot}$. 

\subsection{Integration of the radiative transfer equations}
\label{subsection:fortrancode}
We developed a specific \textsc{fortran} code to solve the radiative transfer 
in the fireball atmosphere described in section \ref{subsection:radiativetransfer}.
We assumed a maximum optical depth $\tauR^\maxrm=1000$ at the base of the 
fireball atmosphere, since this guarantees that the medium is optically 
thick for X-mode photons over the entire energy range $1$--$100$ 
keV. The optical depth varies between $\tauR^\maxrm$ at the base and $\tauR^\minrm=0$ 
at the top and is sampled by a non-uniform, 500-point grid. However, such 
a high value for $\tauR^\maxrm$ implies extremely large values of $\tau_\ord$ 
(at the strongest magnetic fields) and $\tau_\extr$ (at high energies), 
which make the computational time totally unacceptable. For this reason 
we decide to integrate the two radiative transfer equations (\ref{equation:finRTEtauR}) 
for $\tau_{\ord,\extr} \leq 10$ only, taking $n_{\ord,\extr}$ as Planck 
distributions at the local temperature $T$ otherwise. Since, for the chosen 
values of the parameters, the X-mode photosphere lies always at larger $\tauR$ 
with respect to the O-mode one, we partially modify the second of the equations 
(\ref{equation:finRTEtauR}) in the case of $\tau_\extr\leq 10$ and $\tau_\ord>10$ 
taking the number intensity $n_\ord$ that appears at the right-hand side 
as planckian \cite[see also][]{lyub02}.

As discussed in section \ref{subsection:radiativetransfer}, the fireball 
atmosphere is divided into 30 plane-parallel slabs, labelled by the values
of $B$ at their centres. The (dipolar) magnetic field at the pole is taken 
to be $2\times10^{14}$ G, which results in a minimum value of $B\sim10^{13}$ 
G at the magnetic equator for $R_\maxrm=2R_{NS}$, i.e. at the largest radial 
distance reached by the fireball. The angle $\theta_\mathrm{B}$ between 
the stellar $B$-field and the local normal is always $90^\circ$, because, 
along the field lines, the magnetic field $\boldsymbol{B}$ at the patch 
centre is always perpendicular to the local normal. For the sake of simplicity, 
we assume $\boldsymbol{B}$ as constant over each patch, i.e. we neglect 
its variation in both direction and intensity within the slab. It can be 
verified that the error amounts at most to $\Delta B/B\sim 3\%$. For each 
patch the temperature distribution (\ref{equation:temperaturedistrib}) has 
been assumed, with a bolometric temperature $T_\mathrm{b}=10$ keV.

\begin{figure*}
\begin{center}
\includegraphics[width=16cm]{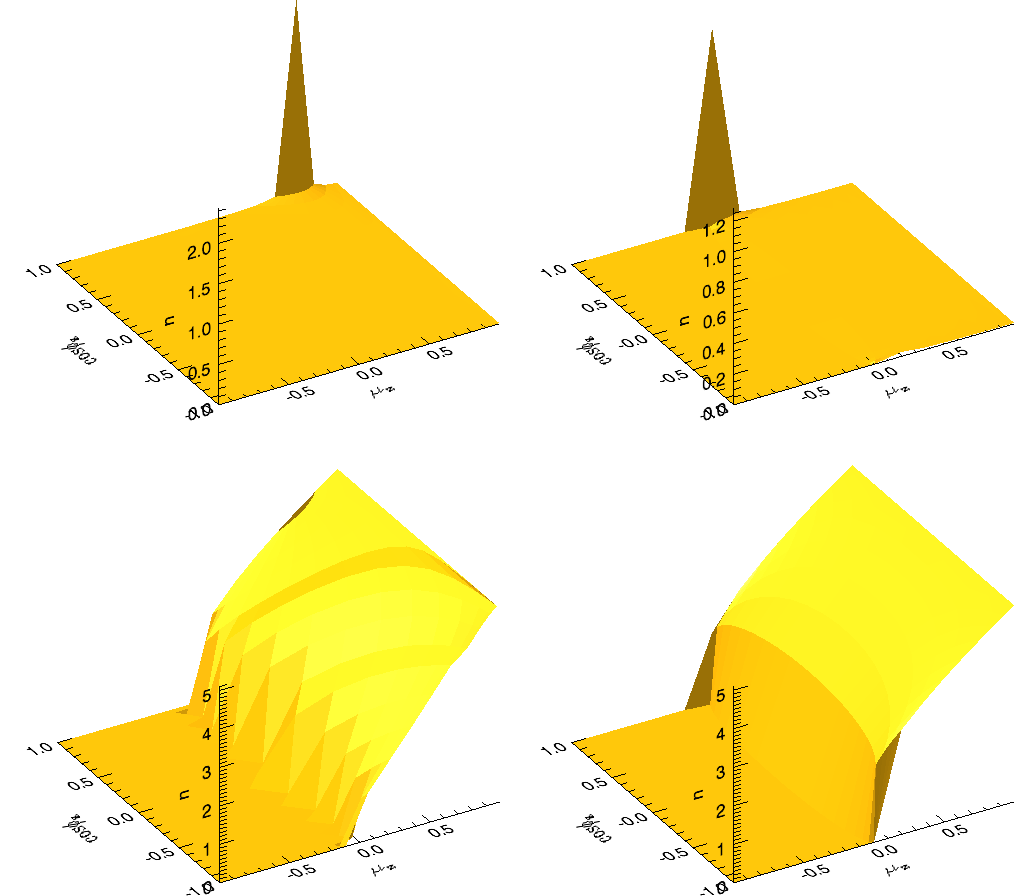}
\caption{Number intensities $n_\ord$ (top row) and $n_\extr$ (bottom row)
of the emerging radiation, plotted as functions of $\mu_z$ and $\cos\phi_z$ 
in the cases of two patches characterized by $\theta_\mathrm{B}=45^\circ$ 
(left-hand column) and $\theta_\mathrm{B}=90^\circ$ (right-hand column); 
the magnetic field intensity is $B=10^{14}$ G.}
\label{figure:angdistr}
\end{center}
\end{figure*}
In order to solve the radiative transfer equations in each slab, the code 
uses a simple $\Lambda$-iteration. In fact, both the equations (\ref{equation:finRTEtauR})
can be written in the compact form
\begin{flalign} \label{equation:lambdaiterRTE}
\frac{\der n_i}{\der\tauR}&=C_in_i-S_i\,\,\,\,\,\,\,\,\,(i=\ord,\,\extr)\,, &
\end{flalign}
where $C_i$ does not depend on $\tauR$ and $S_i$ represents the source term, 
that contains the integrals of $n_i$ over the angles $\theta'_\Bk$ and $\phi'_\Bk$. 
The Sommerfeld radiative condition is imposed at the top of the slab, i.e.
$n_i=0$ for \mbox{$-1\leq \mu_z\leq 0$} at $\tauR=0$. Integration is started 
making an initial guess for $n_i$ and the initial source terms $S^{(0)}_i$ 
are then calculated through a Gauss-Lobatto quadrature, using a 20-point 
grid for both $\mu'_\Bk$ and $\phi'_\Bk$. With these values for the source 
terms, the code integrates the radiative transfer equations using a Runge-Kutta, 
fourth order method, following a set of rays sampled by a 20$\times$20 mesh 
in $-1\leq\mu_\Bk\leq1$ and $0\leq\phi_\Bk\leq2\pi$. The new values of $n_i$ 
are used to calculate the source terms $S^{(1)}_i$ at the next step. The 
iterative method proceeds until the fractional accuracy 
\begin{flalign} \label{equation:fracacc}
\Delta&=\frac{|S_i^{(n)}-S_i^{(n-1)}|}{|S_i^{(n)}|+|S_i^{(n-1)}|} &
\end{flalign}
drops below a given value. The same procedure is repeated for different 
values of the photon energy $\varepsilon$, ranging between 1 and 100 keV 
within an equally-spaced, 30-point grid. 

\begin{figure*}
\begin{center}
\includegraphics[width=16cm]{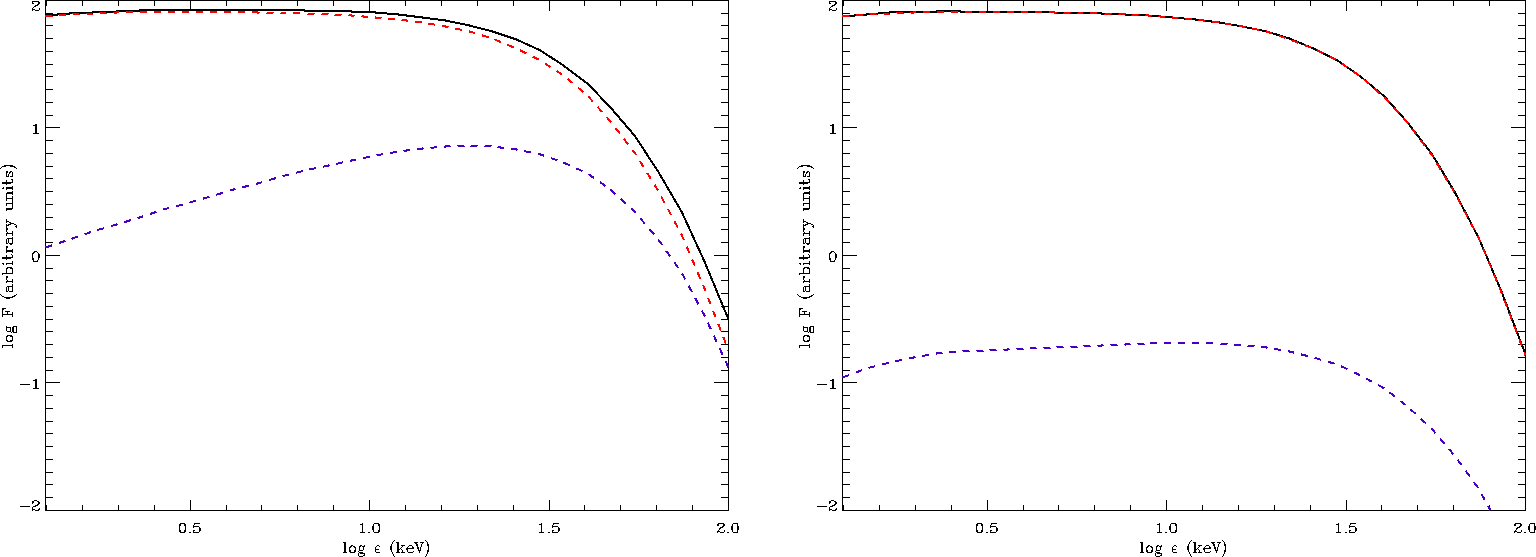}
\caption{Photon number fluxes in the case of two single patches characterized 
by $\theta_\mathrm{B}=90^\circ$, $B=10^{13}$ (left panel) and $10^{14}$ 
G (right panel), plotted as functions of the photon energy. The black, solid 
line represents the total spectrum, while the blue and red dashed lines 
the O-mode and X-mode photon spectra, respectively.}
\label{figure:fluxspectfortran}
\end{center}
\end{figure*}

The code returns the number intensity $n_i$ as a function of the optical 
depth $\tau$ in the slab, the photon energy $\varepsilon$ and the angles 
$\theta_\Bk$ and $\phi_\Bk$ that the photon direction makes with the stellar 
magnetic field at the patch centre. This has been done for 30 selected values 
of the magnetic field intensity (equally spaced in log between $10^{13}$ 
and $2\times10^{14}$ G). Figure \ref{figure:angdistr} presents the angular 
distribution of the emerging radiation ($\tauR=0$) for both ordinary and 
extraordinary photons; for comparison, two different values of $\theta_\mathrm{B}=45^\circ$ 
and $90^\circ$ are shown. In all the panels, the intensities are plotted 
as functions of $\mu_z$ and $\cos\phi_z$, where the latter is the azimuthal 
angle associated to $\theta_z$ (see section \ref{subsection:radiativetransfer}). 
Ordinary photons appear to be beamed along the local magnetic field direction, 
at variance with the extraordinary ones. Figure \ref{figure:fluxspectfortran} 
shows instead the behavior of the total, O-mode and X-mode photon number 
fluxes as functions of the energy for two different patches, with $\theta_\mathrm{B}=90^\circ$ 
and $B=10^{13}$, $10^{14}$ G, respectively. In both the panels it can 
be clearly seen the flattening of the emerging spectrum at low energies, 
due to the fact that X-mode photons with lower energies escape the fireball 
medium from regions with higher temperature \cite[see][]{lyub02}. Moreover, 
the extraordinary photon flux turns out to be well above the ordinary one 
especially for low energies, as expected since the O-mode optical depth 
is much larger than the X-mode one (see section \ref{subsection:radiativetransfer}). 
This difference is enhanced for increasing magnetic field intensities, when 
the X-mode flux dominates over the O-mode one in the entire energy range.

\subsection{Ray-tracing}
\label{subsection:idlcode}
The data generated by the radiative transfer code described above are then
processed in an \textsc{idl} ray-tracer code, in order to simulate the spectra 
of the emitted radiation and the polarization observables at infinity as 
functions of photon energy and rotational phase. The code is discussed in 
detail in \citet[see also \citealt{zanetur06,gonzetal16}]{tavetal15}; given 
that the geometry (see below) is more complicated in the present case, no 
GR corrections are included.

\subsubsection{Geometry}
The neutron star geometry is described in a fixed frame $(X,Y,Z)$ where 
the $Z$ axis is chosen along the line-of-sight (LOS, unit vector $\boldsymbol{\ell}$)
and the $X$ axis is in the plane made by $\boldsymbol{\ell}$ and the star 
spin axis $\boldsymbol{\Omega}$. However it is convenient to introduce also 
a reference frame $(p,q,t)$ which rotates around $\boldsymbol{\Omega}$, 
with the $t$ axis along the star magnetic axis (unit vector $\bdip$). Denoting 
with $\chi$ and $\xi$ the angles that the rotation axis makes with $\boldsymbol{\ell}$ 
and $\bdip$, respectively, the unit vectors that define the $\bdip$ reference 
frame can be expressed in the LOS frame as \cite[see e.g.][]{tavetal15}
\begin{flalign}
\boldsymbol{p} &= (-\sin\chi\sin\xi-\cos\chi\cos\xi\cos\gamma,\cos\xi\sin\gamma, & \nonumber \\
\ &\,\,\,\,\,\,\,\,\, \sin\chi\cos\xi\cos\gamma-\cos\chi\sin\xi) & \nonumber
\end{flalign}
\begin{flalign} 
\boldsymbol{q} &= (-\cos\chi\sin\gamma,-\cos\gamma,\sin\chi\sin\gamma) \nonumber &
\end{flalign}
\begin{flalign} \label{equation:pqt}
\boldsymbol{t} &\equiv\bdip = (\sin\chi\cos\xi-\cos\chi\sin\xi\cos\gamma,\sin\xi\sin\gamma, & \nonumber \\
\ &\,\,\,\,\,\,\,\,\,\,\,\,\,\,\,\,\,\,\,\,\,\,\,\,\,\,\, \cos\eta)\,, &
\end{flalign}
where $\gamma$ is the star rotational phase and $\cos\eta=\cos\chi\cos\xi+
\sin\chi\sin\xi\cos\gamma$ is the cosine of the angle between $\boldsymbol{\ell}$ 
and $\bdip$. 

The position of a generic point on the surface of the fireball is given by
\mbox{$\boldsymbol{r}=r(\sin\Theta\cos\Phi,\sin\Theta\sin\Phi,\cos\Theta)$}
in the LOS reference frame, where the radial distance $r$ is given by the 
relation
\begin{flalign} \label{equation:fieldlines}
r&=R_\maxrm\sin^2\theta\,. &
\end{flalign}
Because $r\geq R_\mathrm{NS}$, the magnetic colatitude $\theta$ is in the 
range $[\theta_\minrm,\theta_\maxrm]$, with
\begin{flalign} \label{equation:thetamin}
\theta_\minrm&=\arcsin\left(\sqrt{\frac{R_\mathrm{NS}}{R_\maxrm}}\right) &
\end{flalign}
and $\theta_\maxrm=\pi-\theta_\minrm$. This holds for each value of the 
magnetic azimuth $\phi$, that ranges between $0$ and $2\pi$ if the fireball 
is assumed to be torus-like, i.e. it fills the entire volume between the 
star surface and the limiting field lines given by equation (\ref{equation:fieldlines}); 
for the sake of conciseness, in the following we will refer to this volume 
as the fireball torus. The polar angles $\theta$ and $\phi$ in the $\bdip$ 
frame are then related to $\Theta$ and $\Phi$ in the LOS frame by
\begin{flalign} \label{equation:thetaphiGRpi}
\cos\theta&=\boldsymbol{r}\cdot\bdip & \nonumber \\
\cos\phi&=\boldsymbol{r}_\perp\cdot\boldsymbol{p}\,, &
\end{flalign}
where
\begin{flalign} \label{equation:rperp}
\boldsymbol{r}_\perp&=\frac{\boldsymbol{r}-(\bdip\cdot\boldsymbol{r})\bdip}
{|\boldsymbol{r}-(\bdip\cdot\boldsymbol{r})\bdip|} &
\end{flalign}
is the (normalized) projection of the position vector $\boldsymbol{r}$ in 
the plane perpendicular to $\bdip$. 

We consider also the case in which the azimuthal extension of the fireball 
is restricted between two given values $\phi_\minrm$ and $\phi_\maxrm$, 
so that the emitting surface consists of a portion of the torus plus the
two constant-$\phi$ cuts. Of course, the magnetic field is perpendicular
to the surface normal also within the two $\phi_\minrm,\phi_\maxrm$ ``sides''
of the torus. For shortness, we will refer hereafter to the case of emission
from the entire torus as ``model a'', while we will call ``model b'' that 
in which emission is limited to the portion of the fireball between $\phi_\minrm$
and $\phi_\maxrm$ (with the emission from the two sides included).

\subsubsection{Fireball visibility} \label{subsubsect:visibility}
\begin{figure}
\begin{center}
\includegraphics[width=8cm]{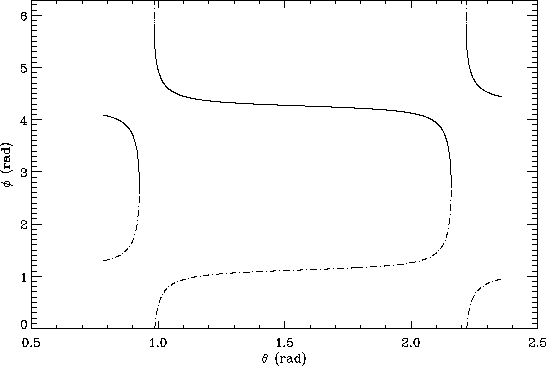}
\caption{Behavior of the first (solid line) and of the second (dash-dotted
line) solutions of equation (\ref{equation:terminatorequation}), plotted 
as functions of the magnetic colatitude $\theta$, for $\chi=60^\circ$, $\xi=30^\circ$
and $\gamma=150^\circ$. Notice that the dash-dotted solution has been reduced
to the $[0,2\pi]$ range.}
\label{figure:terminatorsolution}
\end{center}
\end{figure}
\begin{figure*}
\begin{center}
\includegraphics[width=16cm]{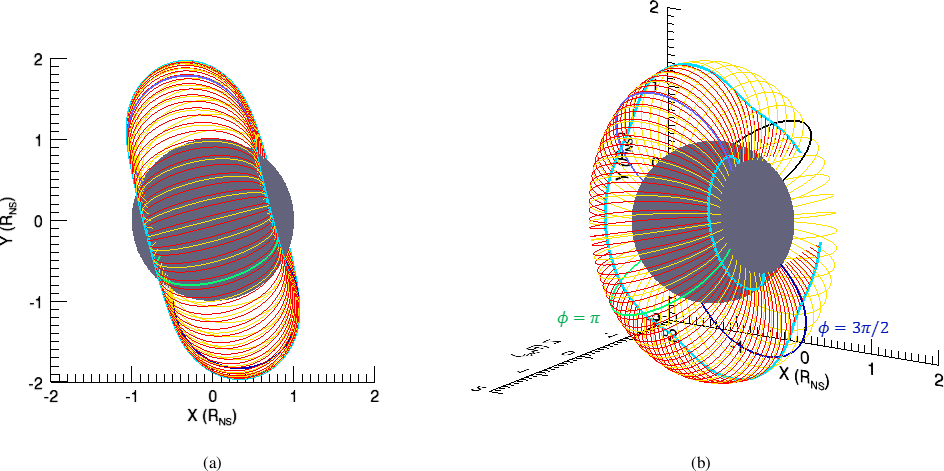}
\caption{Limiting field lines (yellow lines) of a torus-shaped fireball
for $\chi=60^\circ$, $\xi=30^\circ$ and $\gamma=150^\circ$, seen both along 
the LOS ($Z$ axis, left-hand panel) and slightly rotated around the $Y$ 
axis (right-hand panel); the field lines corresponding to $\phi=0^\circ,\,
90^\circ,\,180^\circ,\,270^\circ$ are highlighted in black, violet, green 
and dark blue, respectively. The terminator is marked by the light blue
line, while the region in view of the fireball surface is marked by the 
red lines.}
\label{figure:terminatorexample}
\end{center}
\end{figure*}
To determine the portion of the fireball surface which is in view once the 
values of the geometrical angles $\chi$ and $\xi$ have been fixed, the code 
calculates the terminator of the surface, that is given by the condition
\begin{flalign} \label{equation:terminatorgen}
\boldsymbol{z}\cdot\boldsymbol{\ell}&=0\,, &
\end{flalign}
where $\boldsymbol{z}$ denotes the local surface normal unit
vector. The calculation turns out to be simpler in the $\bdip$ reference
frame; here the LOS unit vector $\boldsymbol{\ell}$ takes the form \mbox{
$\boldsymbol{\ell}=(\sin\eta\cos\delta,\sin\eta\sin\delta,\cos\eta)$}, with
$\delta$ the azimuth of the LOS with respect to $\bdip$, while the components 
of the local normal $\boldsymbol{z}$ are derived in Appendix \ref{appendix:localnormal}. 
In particular, $\delta$ is related to the angles $\chi$, $\xi$ and $\gamma$
through the scalar product
\begin{flalign} \label{equation:angledelta}
\cos\delta&=\boldsymbol{p}\cdot\boldsymbol{\ell}_\perp\,, &
\end{flalign}
that can be calculated in the LOS frame, where the components of the unit
vector $\boldsymbol{p}$ are given by the first of equations (\ref{equation:pqt}),
while those of the normalized projection $\boldsymbol{\ell}_\perp$ of the 
LOS orthogonal to $\bdip$ result
\begin{flalign} \label{equation:lperp}
\boldsymbol{\ell}_\perp&=\frac{1}{\sin\eta}\left(\begin{array}{c}
-b_\mathrm{dip,X}\cos\eta \\ -b_\mathrm{dip,Y}\cos\eta \\ 1-b_\mathrm{dip,Z}\cos\eta
\end{array}\right)\,. &
\end{flalign}
Using equation (\ref{equation:surfacenormalpart}), the condition (\ref{equation:terminatorgen}) 
translates into the following equation,
\begin{flalign} \label{equation:terminatorequation}
\cos(\phi-\delta)&=\frac{3\sin\theta\cos\theta\cos\eta}{\sin\eta(3\cos^2\theta-1)}\,, &
\end{flalign}
which can be solved only if the condition
\begin{flalign} \label{equation:terminatorinequality}
\left|\frac{3\sin\theta\cos\theta\cos\eta}{\sin\eta(3\cos^2\theta-1)}\right| &< 1 &
\end{flalign}
is satisfied (the details of the calculations are discussed in Appendix 
\ref{appendix:terminator}). The solutions of equation (\ref{equation:terminatorequation})
can be written in the simple form
\begin{flalign} \label{equation:terminatorsolution}
&\phi=\arccos A+\delta\,\,\,\cup\,\,\,\phi=(2\pi-\arccos A)+\delta\,, &
\end{flalign}
where $A$ denotes the right-hand side of equation (\ref{equation:terminatorequation})
for $\theta$ belonging to the domain (\ref{equation:terminatordomain}).
Since both $A$ and $\delta$ depend on the values of $\chi$, $\xi$ and the
rotational phase $\gamma$, the shape of the surface terminator will change
for different viewing geometries. As an example, Figure \ref{figure:terminatorsolution}
shows the behavior of the two solutions (\ref{equation:terminatorsolution}) 
for $\chi=60^\circ$, $\xi=30^\circ$ and $\gamma=150^\circ$. It can be clearly 
seen that, for the given values of the viewing angles, the solutions exist 
only within the three intervals $I_1=[\theta_\minrm,a_1]$, $I_2=[a_4,b_1]$ 
and $I_3=[b_4,\theta_\maxrm]$, where $a_1$, $a_4$, $b_1$ and $b_4$ are defined 
in equation (\ref{equation:terminatordomain}). Figure \ref{figure:terminatorexample} 
shows the terminator and the part in view of the fireball surface for the 
same values of $\chi$, $\xi$ and $\gamma$.

Having calculated the polar angles $\theta$ and $\phi$ that characterize 
the terminator in the $\bdip$ reference frame, it remains to determine which 
points on the fireball surface are in view. The code first
calculates the values of colatitude $\theta_\mathrm{term}^{\pm}$ that identify 
the terminator at each given value of the magnetic azimuth $\phi$. This 
is equivalent to draw a horizontal line in the plot of Figure \ref{figure:terminatorsolution},
and find the intersections of this line with the curves. The latter can 
be found inverting equation (\ref{equation:terminatorequation}),
\begin{flalign} \label{equation:thetaterm}
\tan\theta_\mathrm{term}^{\pm}&=\frac{-3\cos\eta\pm\sqrt{9\cos^2\eta+8\cos^2\big(\phi-\delta\big)\sin^2\eta}}{2\cos\big(\phi-\delta\big)\sin\eta}\,. &
\end{flalign}
\begin{figure}
\begin{center}
\includegraphics[width=8cm]{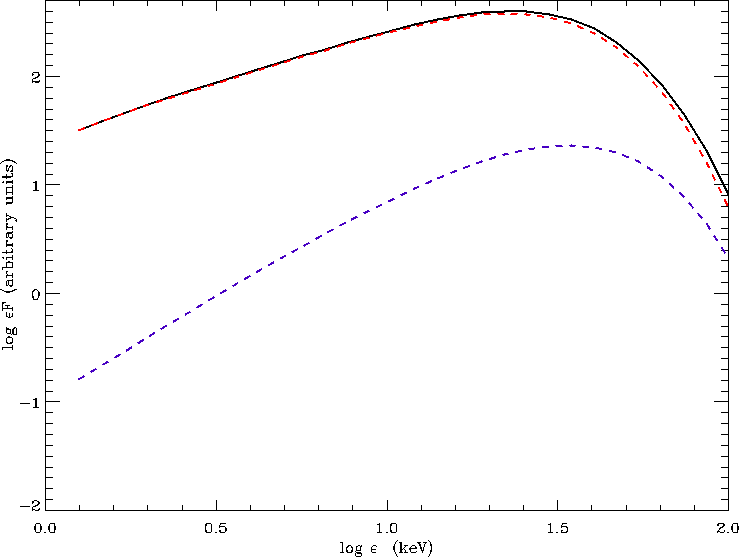}
\caption{Phase-averaged total (black, solid line), ordinary (blue, dashed
line) and extraordinary (red, dashed line) photon spectra obtained from 
the ray-tracer code in the case of model a, for $\chi=60^\circ$ and $\xi=30^\circ$ 
(i.e. the case shown in Figure \ref{figure:terminatorexample}).}
\label{figure:spectidla}
\end{center}
\end{figure} 
Taking e.g. $\phi=\pi$, equation (\ref{equation:thetaterm}) gives two values 
$\theta_\mathrm{term}^{\pm}$ that correspond to the terminator and lie in 
the intervals $I_1$ and $I_2$ defined above. These two terminator points 
are clearly visible in Figure \ref{figure:terminatorexample}, where the 
field line corresponding to $\phi=\pi$ is marked in green; specifically,
it can be seen that the only points along the field line that are in view 
are those included between the two values $\theta_\mathrm{term}^\pm$. On 
the other hand, in the case $\phi=3\pi/2$ Figure \ref{figure:terminatorsolution} 
shows that the two solutions $\theta_\mathrm{term}^{\pm}$ of equation (\ref{equation:thetaterm})
belong to the intervals $I_2$ and $I_3$. In this case, the points in view 
along the field line (see the dark-blue line in Figure \ref{figure:terminatorexample}b) 
are those connecting $\theta_\minrm$ and the smaller of $\theta_\mathrm{term}^\pm$. 
Although this qualitative reasoning has been discussed for two values of 
$\phi$ only, it allows to get a general rule that holds for whatever field 
line, labeled by a generic magnetic azimuth. Given a value of $\phi$, the 
points of the field line in view are those with $\theta$ between the two 
solutions $\theta_\mathrm{term}^{\pm}$ (given in equation \ref{equation:thetaterm}) 
if one of them belong to the interval $I_1$; only the points with colatitude 
within $\theta_\minrm$ and the smaller of $\theta_\mathrm{term}^{\pm}$ will 
be in view otherwise\footnote{For values of the viewing angles such that 
the southern magnetic pole is in view for certain phases, instead of the 
northern one (as shown in Figure \ref{figure:terminatorexample}), a similar 
criterion holds, with the values of the magnetic colatitude $\theta$ replaced 
by $\pi-\theta$ to account for the North-South symmetry.}.

In the case in which the fireball is limited azimuthally between $\phi_\minrm$ 
and $\phi_\maxrm$, a different visibility condition has to be imposed for
the two constant-$\phi$ ``sides''. Actually, since we assumed that these 
two surfaces are planar, the problem can be simply solved checking the sign 
of $\boldsymbol{\ell}\cdot\boldsymbol{h}$, with $\boldsymbol{h}$ the outgoing 
normal to the constant-$\phi$ slice. If this scalar product is positive 
then all the points of the slice are in view. In particular, in the $\bdip$ 
reference frame, the unit vector $\boldsymbol{h}$ for each of the two slices 
can be expressed as 
\begin{flalign} \label{equation:wallnormal}
\boldsymbol{h}&=\pm\,\boldsymbol{m}\times\bdip\,, &
\end{flalign}
where the $+$ ($-$) sign corresponds to $\bar{\phi}=\phi_\minrm$ ($\phi_\maxrm$)
and $\boldsymbol{m}$ is the (unit) position vector of the generic point 
in the plane $\phi=\bar{\phi}$ (see the first equality in equation \ref{equation:nthetaphi}). 
By choosing, for the sake of simplicity, $\theta=\pi/2$, one readily obtains
\begin{flalign} \label{equation:hminhmax}
&\boldsymbol{h}=\left(\begin{array}{c}
\pm\sin\bar{\phi} \\ \mp\cos\bar{\phi} \\ 0
\end{array}\right)\,. &
\end{flalign}
By deriving the components $(h_X,h_Y,h_Z)$ in the LOS frame (see Appendix 
\ref{appendix:changeofbasis}), the visibility condition can be then written
as \mbox{$h_Z>0$}.
\begin{figure*}
\begin{center}
\includegraphics[width=17.5cm]{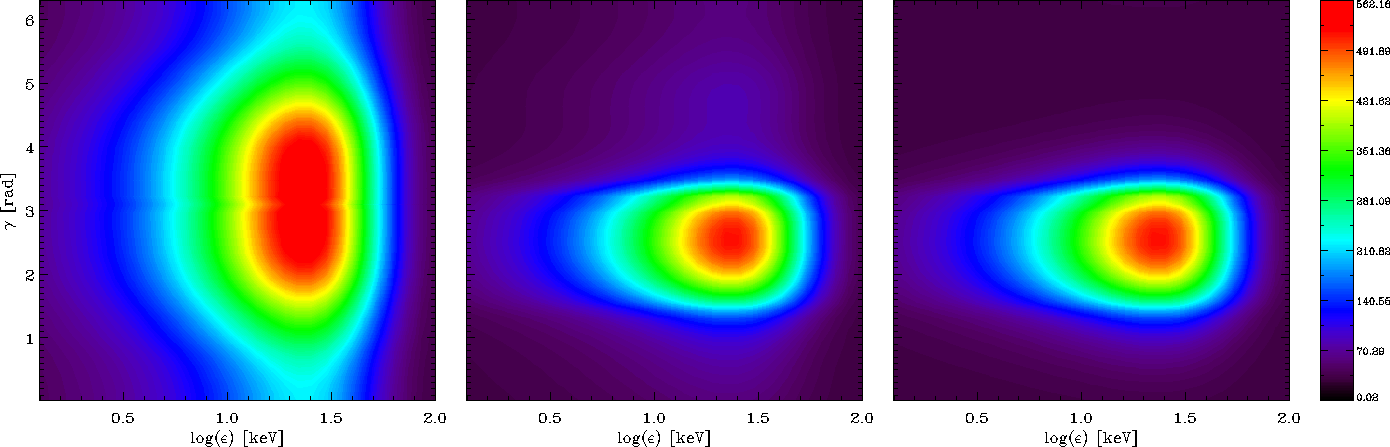}
\caption{Total intensity (arbitrary units), plotted as a function of the photon 
energy $\varepsilon$ and the rotational phase $\gamma$, in the case of model 
a (left-hand panel) and model b with $\phi_\minrm=100^\circ$ and $\phi_\maxrm=190^\circ$ 
(central panel). The right-hand panel shows the same case as the central 
one, but without including emission from the planar slices at the sides. 
As in Figure \ref{figure:spectidla}, it is $\chi=60^\circ$ and $\xi=30^\circ$.}
\label{figure:phaseresolved}
\end{center}
\end{figure*}

Finally, one needs to exclude, from the part in view of both the torus and
the slices found with the methods discussed above, the region covered by 
the star projection in the plane of the sky. This can be easily performed 
transforming the coordinates of the position vector $\boldsymbol{m}$ of 
each point in view from the $\bdip$ frame (see equation \ref{equation:nthetaphi})
to the LOS one. Hence, the points covered by the star are identified by 
the condition
\begin{flalign} \label{equation:starcovered}
&m^2_X+m^2_Y<1\,\,\,\cap\,\,\,m_Z<1 &
\end{flalign}
and can be excluded. The result of this operation is evident in Figure \ref{figure:terminatorexample}b.

\subsubsection{Photon flux and polarization observables} \label{subsubsec:stokesflux}
In the ray-tracer code the entire fireball surface is divided into a $100$$\times$$100$ 
angular mesh in $(\Theta,\Phi)$, where \mbox{$0<\Theta<\pi$} and \mbox{$0<\Phi<2\pi$}, 
while each of the two sides at $\phi_\minrm$ and $\phi_\maxrm$ (if present) 
is divided through a $(r,\theta)$ grid with \mbox{$\theta_\minrm\leq\theta
\leq\theta_\maxrm$} and \mbox{$R_\mathrm{NS}\leq r\leq R_\maxrm\sin^2\theta$} 
(in the actual calculation a $50\times10$ mesh was used). Once the visible 
part of the fireball is known, any given patch is characterized by the value 
of the strength $B$ of the magnetic field and by the polar angles $\theta_\Bk$ 
and $\phi_\Bk$ that the ray which reaches the observer (i.e. which propagates 
along the LOS) makes with the local $B$-field direction. Then, the code 
calculates the total photon flux and the polarization observables summing 
the contributions from the single surface patches. In this respect, it is 
important to notice that the surface element takes different forms according
to which part of the fireball (the toroidal surface or the ``sides'') is 
considered. In particular, for the toroidal surface, one has in the LOS
reference frame
\begin{flalign} \label{equation:dA1}
\der A_1&=r^2\sin\Theta\der\Theta\der\Phi=R^2_\maxrm\sin^4\theta\sin\Theta\der\Theta\der\Phi\,, &
\end{flalign}
where equation (\ref{equation:fieldlines}) has been used and $\sin^4\theta$
can be written as a function of $\Theta$ and $\Phi$ using the first of equations 
(\ref{equation:thetaphiGRpi}). Instead, for the two azimuthal slices, it
is convenient to express the surface element in the $\bdip$ frame, obtaining
\begin{flalign} \label{equation:dA2}
\der A_2&=r\der r\der\theta\,. &
\end{flalign}
Therefore, the total photon flux $F_\mathrm{I}$ will be the sum of the two 
contributions $F_{I,1}$ and $F_{I,2}$, with
\begin{flalign} \label{equation:FI1}
F_{\mathrm{I},j}&=F_{\mathrm{O},j}+F_{\mathrm{X},j}=\int\big(n_\ord+n_\extr\big)P_j\der A_j\,; &
\end{flalign}
here the symbol $P_j$ represents the projection factor,
\begin{flalign} \label{equation:projfactor}
P_1&=\boldsymbol{z}\cdot\boldsymbol{\ell} \ \ \ \ \ \ \ \ \ \ P_2=\boldsymbol{h}\cdot\boldsymbol{\ell}\,, &
\end{flalign}
where $\boldsymbol{z}$ and $\boldsymbol{h}$ are defined by equations (\ref{equation:surfacenormalpart})
and (\ref{equation:hminhmax}), respectively.

\begin{figure*}
\begin{center}
\includegraphics[width=15cm]{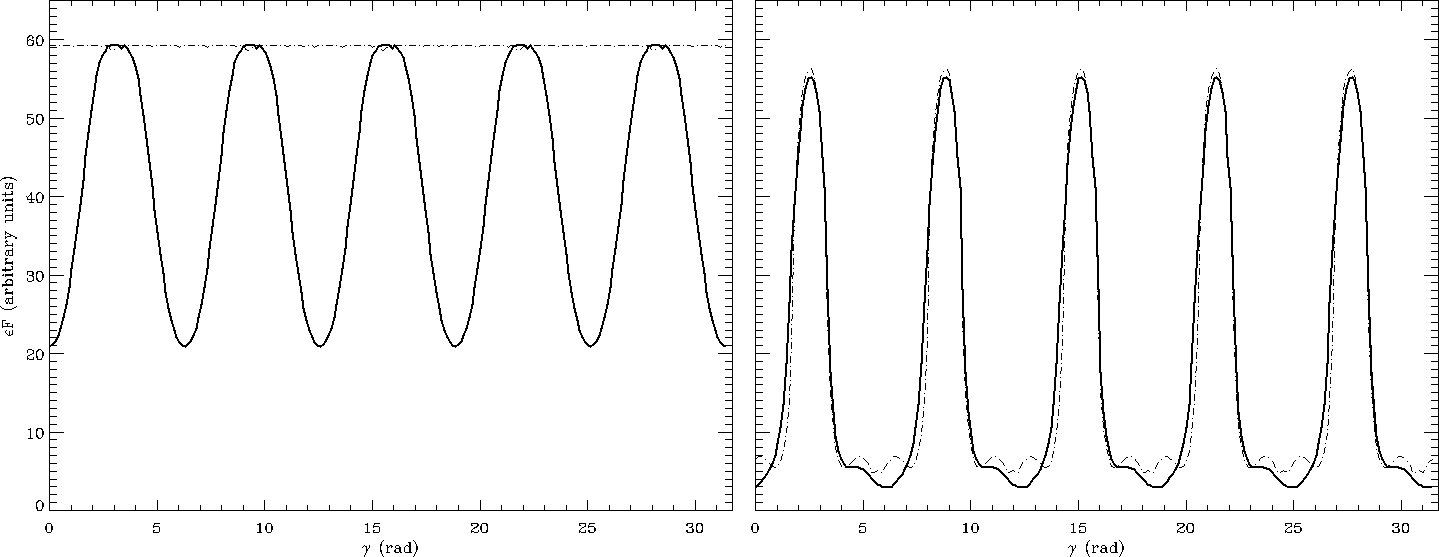}
\caption{Pulse profile of the radiation emitted from the fireball in the 
case of model a (left-hand panel) and model b (with $\phi_\minrm=100^\circ$ 
and $\phi_\maxrm=190^\circ$, right-hand panel), calculated in the $10$--$50$ 
keV energy range. Here the solid lines refers to the case $\chi=60^\circ$, 
$\xi=30^\circ$ and the dash-dotted lines to the case $\chi=90^\circ$, $\xi=0^\circ$.}
\label{figure:pulseprofile}
\end{center}
\end{figure*}
\begin{figure*}
\begin{center}
\includegraphics[width=12cm]{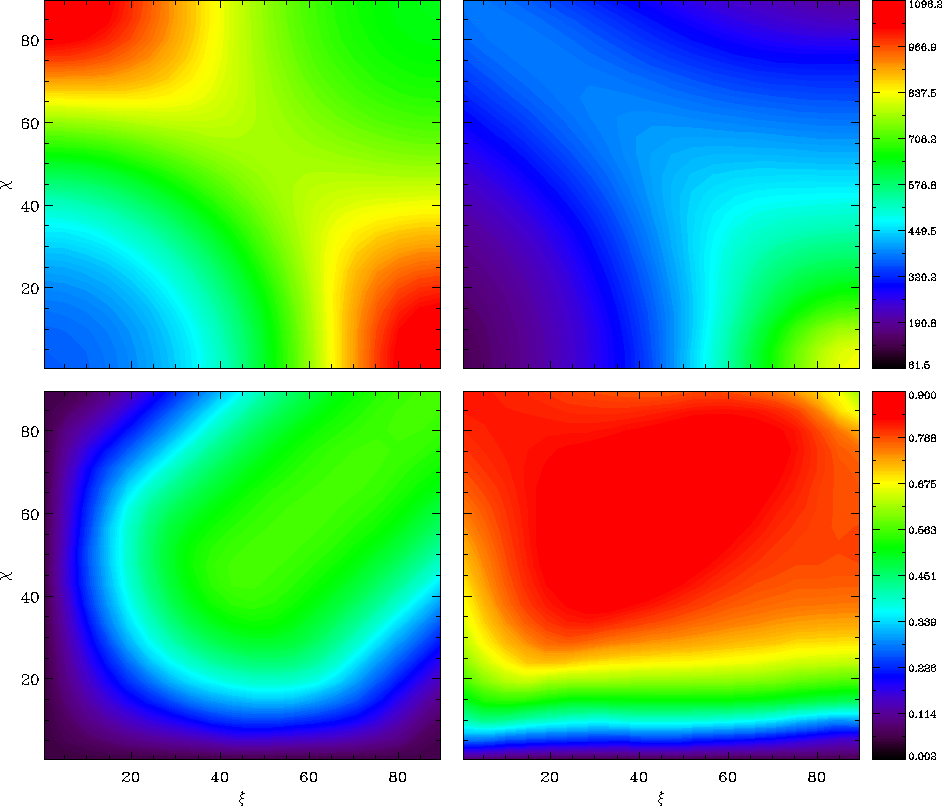}
\caption{Integrated flux (top row) and pulsed fraction (bottom row) of the 
radiation emitted from the fireball in the case of model a (left column) 
and model b (with $\phi_\minrm=100^\circ$ and $\phi_\maxrm=190^\circ$, right 
column), calculated in the $10$--$50$ keV energy range and plotted as functions
of the angles $\chi$ and $\xi$.}
\label{figure:lightflux}
\end{center}
\end{figure*}
In order to account for the effects of quantum electro-dynamics (QED) on
the polarization of the emitted radiation, we use the same simplifying approach 
discussed in \citet[see also \citealt{gonzetal16}]{tavetal15}. Labelling 
$\ell_\mathrm{A}$ and $\ell_\mathrm{B}$ the scale-lengths along which the 
electric field of each photon and the stellar magnetic field evolve, respectively, 
we divide the region in which photons propagate into two zones: the adiabatic 
region (characterized by $\ell_\mathrm{A}\ll\ell_\mathrm{B}$) and the external 
one (where $\ell_\mathrm{A}\gg\ell_\mathrm{B}$), sharply separated at the 
adiabatic radius $r_\mathrm{a}$, defined by the equality $\ell_\mathrm{A}=\ell_\mathrm{B}$. 
According to this approximation, QED effects ensure that the polarization 
vector of each photon istantaneously adapts to the local magnetic field 
direction up to $r_\mathrm{a}$, allowing photons to maintain their initial 
polarization mode. At $r=r_\mathrm{a}$ we assume that the photon electric 
field direction freezes, i.e. it remains oriented in the same direction 
assumed at the adiabatic radius up to the observer. In this way, in order
to reconstruct the polarization properties of the radiation at infinity, 
the single photon Stokes parameters should be rotated, before to be summed
together, by twice the angle $\alpha$ between the local frame $(x,y,z)$ 
of each photon (with the $z$ axis along the LOS and the $x$ axis perpendicular 
to the $\boldsymbol{\ell},\boldsymbol{B}$ plane at $r_\mathrm{a}$) and the 
fixed frame $(u,v,w)$ of the polarimeter \cite[with the $w$ axis also along 
the LOS and $u$, $v$ a generic pair\footnote{Here, as in \citet{tavetal15}, 
we take the $u$ axis in the plane made by the LOS and the star rotation 
axis $\boldsymbol{\Omega}$.} of orthogonal axes perpendicular to $w$, see][for 
further details]{tavetal15}. Generalizing, then, the discrete sum to a continuos 
photon distribution, we can define the ``fluxes'' of Stokes parameters as
\begin{flalign} \label{equation:FQFU}
F_\mathrm{Q}&=\displaystyle{\sum\limits_{j=1}^2\int\big(n_\extr-n_\ord\big)\cos(2\alpha)P_j\der A_j} & \nonumber \\
\ &\ & \nonumber \\
F_\mathrm{U}&=\displaystyle{\sum\limits_{j=1}^2\int\big(n_\ord-n_\extr\big)\sin(2\alpha)P_j\der A_j}\,, &
\end{flalign}
where the values taken by $\alpha$ depend on the geometrical angles $\chi$
and $\xi$, the rotational phase $\gamma$ and the polar angles $\Theta$ and
$\Phi$ that identify the photon emission points in the LOS frame. The Stokes 
parameter fluxes are finally used to calculate the polarization observables 
$\Pi_\mathrm{L}$ and $\chi_\mathrm{p}$, i.e. the linear polarization fraction 
and the polarization angle, defined as
\begin{flalign} \label{equation:polarizationobservables}
\Pi_\mathrm{L}&=\dfrac{\sqrt{F^2_\mathrm{Q}+F^2_\mathrm{U}}}{F_\mathrm{I}} & \nonumber \\
\chi_\mathrm{p}&=\dfrac{1}{2}\arctan\left(\dfrac{F_\mathrm{U}}{F_\mathrm{Q}}\right)\,. &
\end{flalign}

The intensities $n_{\ord,\extr}$ entering equations (\ref{equation:FI1})
and (\ref{equation:FQFU}) are taken from the set of models we have computed 
in advance (see section \ref{subsection:fortrancode}). Trilinear interpolation 
has been used to obtain the intensities at the required values of $B$ and 
of the polar angles $\theta_\Bk$, $\phi_\Bk$. In this respect it is useful 
to notice that the latter are related to the polar angles of the $\bdip$ 
frame $\theta$, $\phi$ (that can be in turn related to the LOS polar angles 
$\Theta$, $\Phi$ using equations \ref{equation:thetaphiGRpi}) by
\begin{flalign}
\mu_\Bk\equiv\cos\theta_\Bk&=\boldsymbol{B}\cdot\boldsymbol{\ell} & \nonumber
\end{flalign}
\begin{flalign} \label{equation:thetaBk}
\cos\phi_\Bk&= \bar{\boldsymbol{x}}\cdot\bar{\boldsymbol{\ell}}\,, &
\end{flalign}
where $\bar{\boldsymbol{x}}$ and $\bar{\boldsymbol{\ell}}$ are the unit 
vectors along the projection of the local normal $\boldsymbol{z}$ (see Appendix 
\ref{appendix:localnormal}) and $\boldsymbol{\ell}$ in the plane orthogonal 
to $\boldsymbol{B}$, respectively\footnote{For our particular choice, the
unit vector $\bar{\boldsymbol{x}}$ coincides with the local surface normal.}.

\section{Results} \label{section:simulationsresults}
\begin{figure}
\begin{center}
\includegraphics[width=8cm]{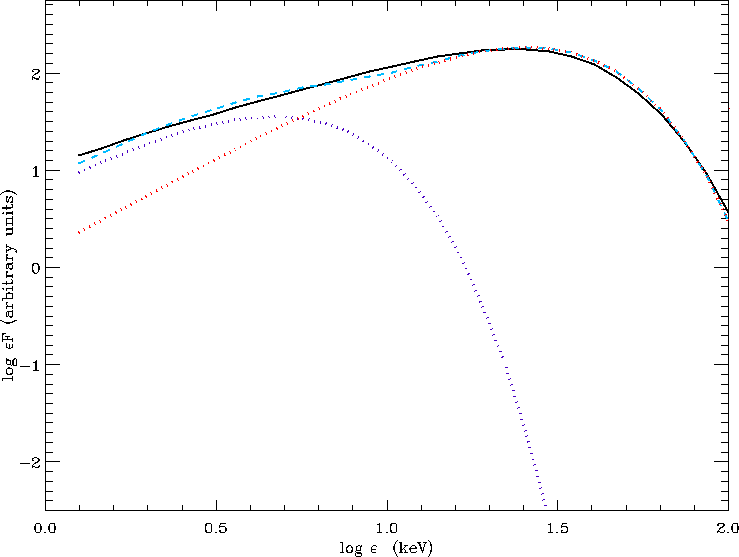}
\caption{Phase-averaged photon spectrum (black solid line) of the radiation
emitted in the case of model b (with $\phi_\minrm=100^\circ$ and $\phi_\maxrm=190^\circ$), 
fitted by the superposition of two blackbody components for $\chi=60^\circ$ 
and $\xi=30^\circ$. The fitting function (\ref{equation:fittingfunction}) 
is marked by the light-blue, dashed line, and the single blackbody components 
at temperature $T_1$ (blue, dotted line) and $T_2$ (red, dotted line) are 
also shown. The results of the fit for different configurations are reported 
in Table \ref{table:fit1d}. }
\label{figure:spectidlb}
\end{center}
\end{figure}
The intensities calculated in the radiative transfer code and processed
in the ray-tracer are then used to obtain the simulated spectra and the 
polarization observables (both phase-resolved and phase-averaged) as observed 
at infinity.
\begin{table}
\begin{center}
\begin{tabular}{lccccc}
\hline
Emitting region & $\chi$ & $\xi$ & $T_1$ (keV) & $T_2$ (keV) & $A_2/A_1$ \\
\hline
Model a & $60^\circ$ & $30^\circ$ & $1.67$ & $9.11$ & $0.0315$ \\
\hline
Model b & $60^\circ$ & $30^\circ$ & $1.67$ & $9.12$ & $0.0315$ \\
\hline
Model a & $90^\circ$ & $0^\circ$ & $1.67$ & $9.18$ & $0.0315$ \\
\hline
Model b & $90^\circ$ & $0^\circ$ & $1.67$ & $9.14$ & $0.0315$ \\
\hline
\end{tabular}
\caption{Results of the fits of the phase-averaged total spectrum using 
the function given by equation (\ref{equation:fittingfunction}), for models
a and b (the latter characterized by $\phi_\minrm=100^\circ$ and $\phi_\maxrm=190^\circ$)
and  for different values of the viewing angles $\chi$ and $\xi$ (the case
illustrated in Figure \ref{figure:spectidlb} is reported in the second row).}
\label{table:fit1d}
\end{center}
\end{table}

\subsection{Phase-averaged and phase-resolved spectra}
\label{subsection:spectra}
\begin{figure*}
\begin{center}
\includegraphics[width=12cm]{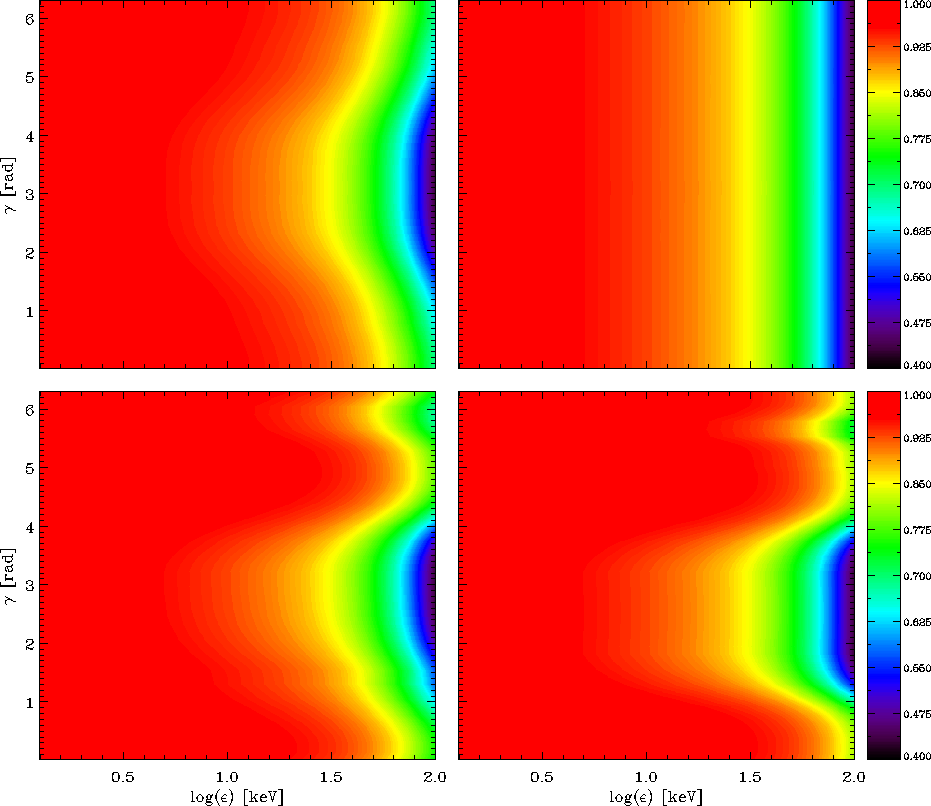}
\caption{Contour plots of the linear polarization degree $\Pi_\mathrm{L}$
as a function of the photon energy $\varepsilon$ and the rotational phase
$\gamma$ for $\chi=60^\circ$, $\xi=30^\circ$ (left-hand column) and $\chi=90^\circ$,
$\xi=0^\circ$ (right-hand column). The top row refers to model a, the bottom 
row to model b (with $\phi_\minrm=100^\circ$ and $\phi_\maxrm=190^\circ$).}
\label{figure:PiLphres}
\end{center}
\end{figure*}
Figure \ref{figure:spectidla} shows the phase-averaged total spectrum of 
the radiation emitted by the entire torus-shaped fireball (model a) for 
$\chi=60^\circ$ and $\xi=30^\circ$ (see Figure \ref{figure:terminatorexample}), 
compared to those of the ordinary and extraordinary components, separately. 
The spectrum of the extraordinary photons appears nearly superimposed to 
the total spectrum at least at lower energies, as already noticed in the 
case of a single patch (see Figure \ref{figure:fluxspectfortran}), while
ordinary photons give a small contribution only above 30 keV. This confirms 
that, although the distributions of the O- and X-mode photons at the base
of the atmosphere are the same (as mentioned in section \ref{subsection:fortrancode}), 
the radiation collected by an observer at infinity is expected to be polarized 
essentially in the extraordinary mode. Conversely, the flux of the collected
ordinary photons is expected to be much lower than for the extraordinary 
ones, the ratio $F_\mathrm{O}/F_\mathrm{X}$ ranging between $\sim 5\times10^{-3}$ 
and $0.3$ across the entire $1-100$ keV energy range. 

The phase-resolved, total spectrum is instead shown in Figure \ref{figure:phaseresolved}, 
for $\chi=60^\circ$, $\xi=30^\circ$ and in the cases of model a (i.e. the 
same case of Figure \ref{figure:spectidla}) and model b with $\phi_\minrm=100^\circ$ 
and $\phi_\maxrm=190^\circ$ (see the discussion in section \ref{subsection:idlcode}). 
For the sake of completeness, we also included the case in which the emission 
from the planar slices at $\phi_\minrm$ and $\phi_\maxrm$ is neglected. 
It is interesting to notice how our simple 
fireball model is indeed able to reproduce a single-pulse flux profile. In this respect, Figure \ref{figure:pulseprofile} 
shows the light curves obtained in the case of model a and model b (with 
emission from the planar slices at $\phi_\minrm$ and $\phi_\maxrm$ included), 
for two different sets of geometrical angles. In particular, from 
the left-hand panel it can be clearly seen that pulsations can be obtained,
for some favourable viewing geometries, regardless of whether the emitting 
region is limited or not. This can be explained with the intrinsic anisotropy 
of the fireball radiation pattern. In fact, as discussed in section \ref{subsection:fortrancode},
the photon flux depends on the angle the LOS makes with the local fireball 
normal (see e.g. Figure \ref{figure:angdistr}). Figure \ref{figure:lightflux}, 
which illustrates the integrated flux and pulsed fraction in the 
$10$--$50$ keV energy range as functions of the angle $\chi$ and $\xi$, 
further confirms this expectation, showing a wide range of variation for 
the flux in the $\chi$--$\xi$ plane also in the case of model a, where emission
comes from the entire torus. Specifically, it attains its maximum value 
when the equatorial regions of the fireball enter into view, while it has 
a minimum when the star is observed along the magnetic axis. On the other 
hand, the pulsed fraction appears to be much more influenced by the dimension 
of the emitting region of the torus. Both Figures \ref{figure:pulseprofile} 
and \ref{figure:lightflux} show that, for equal $\chi$ and $\xi$, the maximum 
values of pulsed fraction are attained in the case of model b, where the 
emitting region is hidden to the observer at certain values 
of the rotational phase.
\begin{figure}
\begin{center}
\includegraphics[width=8cm]{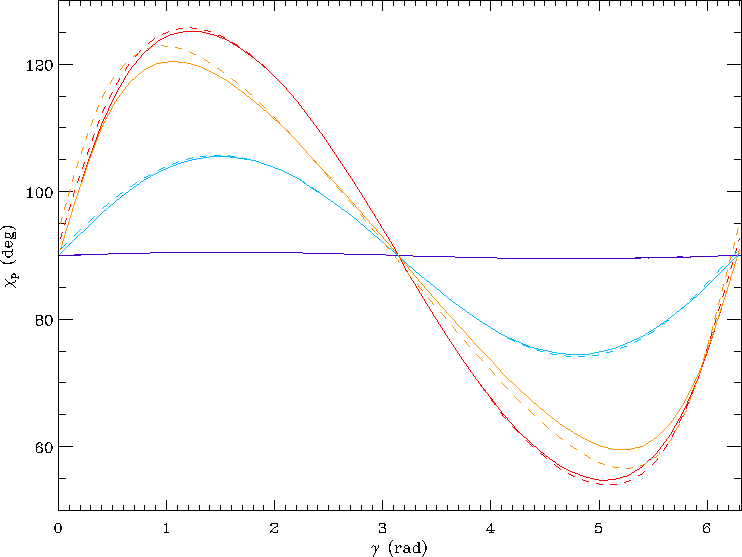}
\caption{Phase-resolved behavior of the polarization angle $\chi_\mathrm{p}$ 
averaged over the entire $1-100$ keV energy range, plotted in the cases 
of model a (solid lines) and model b with $\phi_\minrm=100^\circ$ and $\phi_\maxrm=190^\circ$
(dashed lines) for different viewing geometries: $\chi=20^\circ$, $\xi=10^\circ$ 
(orange), $\chi=60^\circ$, $\xi=30^\circ$ (red), $\chi=75^\circ$, $\xi=15^\circ$ 
(light-blue) and $\chi=90^\circ$, $\xi=0^\circ$ (violet).}
\label{figure:chipolEnAv}
\end{center}
\end{figure}

Observations of the ``burst forest'' emitted by SGR 1900+14 in 2006 \cite[see][]{isretal08}
suggest that the spectrum of the intermediate flares (and of normal bursts 
too) is thermal and well reproduced by the superposition of two blackbodies.
Although comparing results from our simplified model with observations is 
premature, we nevertheless attempted to fit the phase-averaged spectra for
different values of the geometrical angles $\chi$ and $\xi$ with two blackbody
distributions at temperatures $T_1$ and $T_2$,
\begin{flalign} \label{equation:fittingfunction}
f(\varepsilon)&=A_1\varepsilon^3\left(\frac{1}{\exp(\varepsilon/kT_1)-1}+\frac{A_2/A_1}{\exp(\varepsilon/kT_2)-1}\right)\,, &
\end{flalign}
where the normalizations $A_1$ and $A_2$ are related to the emitting areas.
As an example, the total emerging spectrum together with the best fit and 
the single components are shown in Figure \ref{figure:spectidlb}, for $\chi=60^\circ$ 
and $\xi=30^\circ$ and in the case of model b (with $\phi_\minrm=100^\circ$
and $\phi_\maxrm=190^\circ$). Table \ref{table:fit1d} summarizes the fit 
parameters obtained for some selected cases. Actually, there is no significant 
difference switching from model a to model b, and the situation does not 
appreciably change for different viewing geometries. In general, it turns 
out that a representation of the model spectra in terms of two blackbodies 
is satisfactory. With our choice of the temperature distribution (see sections 
\ref{subsection:radiativetransfer} and \ref{subsection:fortrancode}), $T_1$ 
and $T_2$ are $\sim 2$ and $\sim 9$ keV, respectively. Furthermore, the 
ratio $A_2/A_1$ between the emitting areas of the harder and the softer 
components is about $3$\% in all the cases considered, translating in a 
ratio $\sim 0.18$ between the respective blackbody radii. Averaging over 
a number of different events observed during the burst forest of SGR 1900+14, 
\citet[see also reference therein]{isretal08} obtained $T_\mathrm{h}=9.0\pm0.3$ 
keV, $T_\mathrm{s}=4.8\pm0.3$ keV and a ratio $R_\mathrm{h}/R_\mathrm{s}\sim 0.19\pm 
0.03$. Despite the fact that we considered only a single model (with bolometric 
temperature $T_\mathrm{b}=10$ keV and azimuthal extension of the fireball 
$\phi_\maxrm-\phi_\minrm=90^\circ$), theoretical predictions appear to be 
in broad agreement with the observations, although the temperature of the 
softer component turns out to be somehow lower than the observed one. In 
this picture, both the blackbody components required to reproduce the observed 
spectra are essentially made by extraordinary photons (the ordinary ones 
contributing only at high energies, see Figure \ref{figure:spectidla}). 
This appears at variance with the original suggestion by \citet{isretal08} 
that the two components originate from the O- and X-mode photospheres.

\subsection{Polarization observables}
\label{subsection:polarizationobs}
\begin{figure*}
\begin{center}
\includegraphics[width=12cm]{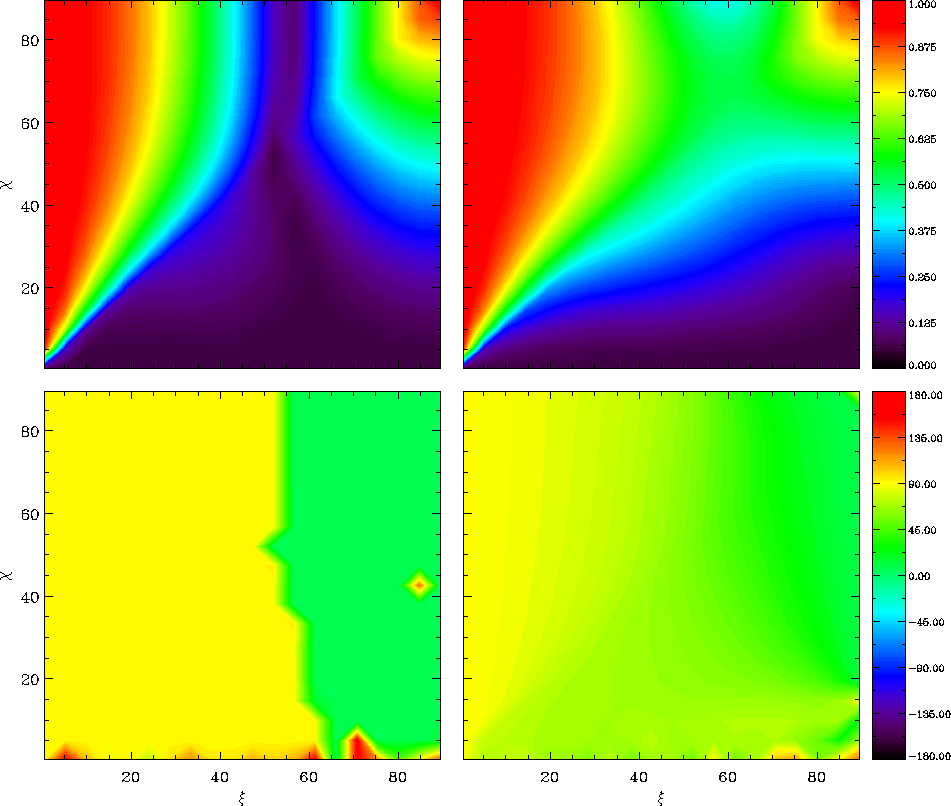}
\caption{Phase-averaged linear polarization fraction $\Pi_\mathrm{L}$ (top 
row) and polarization angle $\chi_\mathrm{p}$ (bottom row) calculated in 
the $10$--$50$ keV energy range and plotted as functions of the angles $\chi$ 
and $\xi$. The left-hand column refers to model a, the right-hand column 
to model b (with $\phi_\minrm=100^\circ$ and $\phi_\maxrm=190^\circ$).}
\label{figure:polobsPhAv}
\end{center}
\end{figure*}
The observed polarization signal, computed taking into account both vacuum 
polarization and the geometrical effects due to the magnetic field topology 
\cite[see][for further details]{tavetal15}, confirms that, according to 
our model, a high degree of polarization is expected for the radiation 
collected from a magnetar flare.

The behavior of the observed linear polarization fraction, as a function 
of the rotational phase and photon energy, is shown in Figure \ref{figure:PiLphres}
for both models a and b and different viewing geometries. In all the cases
considered, $\Pi_\mathrm{L}$ attains a value generally higher than 80\% 
for photon energies between $1$--$50$ keV. However, radiation appears to 
be more polarized at lower energies ($\Pi_\mathrm{L}>90\%$ for $\varepsilon\sim 1-30$ 
keV), while the contrary happens above 50 keV, where the polarization degree 
decreases up to $\sim 40\%$. This behavior, that seems to be opposite to 
that one would expect considering the dependence of the adiabatic radius 
$r_\mathrm{a}$ on the photon energy \cite[see e.g.][]{tavetal15}, can be 
explained looking at the intensity distributions of the ordinary and extraordinary 
photons (see e.g. Figure \ref{figure:spectidla}). In fact, as discussed 
above, the ratio $F_\ord/F_\extr$ increases with the photon energy, justifying 
the substantial decrease of $\Pi_\mathrm{L}$ at higher energies. This is 
primarily due to the contributions from the patches characterized by smaller 
magnetic field intensities, for which the O-mode photon flux is larger and 
comparable to the X-mode one at higher energies (see the left panel of Figure 
\ref{figure:fluxspectfortran}). Moreover, the bottom row of Figure \ref{figure:PiLphres} 
shows that a higher degree of polarization is expected when the emission 
from the planar sides which limit the fireball is considered. Although the 
difference is actually modest, comparing the bottom-left panel of Figure 
\ref{figure:PiLphres} with the rightmost panels of Figure \ref{figure:phaseresolved}, 
it appears that the radiation coming from the limiting slices is in general 
more polarized than that emitted from the remaining part of the torus.

Figure \ref{figure:chipolEnAv} shows the behavior of the polarization angle 
$\chi_\mathrm{p}$ as a function of the rotational phase for both models 
a and b and different values of the viewing angles $\chi$ and $\xi$. Since, 
as noted by \citet{tavetal14,tavetal15}, the polarization angle is essentially 
constant with the photon energy, in this plot we averaged $\chi_\mathrm{p}$ 
over the entire $1-100$ keV energy range. As in the case of surface emission
from a neutron star \cite[][see also \citealt{fd11,tavetal14}]{tavetal15}, 
the polarization angle oscillates with the rotational phase around a value 
of $90^\circ$, with different amplitudes according to the different values 
of $\chi$ and $\xi$. This is the expected behavior for photons mainly polarized 
in the extraordinary mode, as already noticed in section \ref{subsection:spectra}, 
with the choice of the polarimeter reference frame made in section \ref{subsection:idlcode} 
\cite[see also][]{tavetal15}. Very small differences are visible between 
model a (solid lines) and model b (dashed lines) in Figure \ref{figure:chipolEnAv}. 

As in the case of phase-resolved simulations, also the contour plots in 
Figure \ref{figure:polobsPhAv}, that represent the phase-averaged polarization
observables as functions of the angles $\chi$ and $\xi$ in the $10$--$50$
keV energy range, show an overall increase of the polarization fraction
moving from model a to model b. The large depolarization that is visible
in the top row for certain values of the viewing angles is typical of the
dipolar topology of the stellar magnetic field, and it is due to the geometrical 
effect of rotation of the Stokes parameters from the local frame of each 
photon to the polarimeter frame (see section \ref{subsection:idlcode}). 
However, both the patterns of the polarization fraction for model a and 
b are compatible with radiation highly polarized at the emission, as already 
noticed by \cite{gonzetal16} and \cite{mignetal17} in the case of blackbody 
or atmospherical emission from the star. Finally, also the behavior of the
phase-averaged polarization angle is that expected for radiation characterized
by an excess of extraordinary photons.

\section{Discussion and conclusions}
\label{section:discussion}

We have revisited the problem of modeling the spectral and polarization
properties of the radiation emitted during magnetar flares in the context 
of the trapped-fireball model \cite[see][]{td95,td01}. Our code integrates 
the radiative transfer equations for both ordinary and extraordinary photons 
in the fireball atmospheric layer, divided in a number of different patches.
This model generalizes the approach presented by \citet{lyub02}, who treated 
the problem in the one dimensional approximation by solving the radiative transfer 
equation for X-mode photons only, and in the case of $\boldsymbol{B}$ parallel 
to the patch normal. The outputs of the radiative transfer code have been 
then reprocessed through a ray-tracing code (see section \ref{subsection:idlcode}), 
in order to obtain the spectra and the polarization observable distributions 
as measured by a distant observer. Radiation has been assumed to come from 
either the entire torus-like fireball or a portion limited in the azimuthal 
direction between two values $\phi_\minrm$ and $\phi_\maxrm$. In the latter 
case, also emission from the two planar slices at the boundaries has been 
accounted for. The contributions from the patches which enter into view 
are finally summed together, providing the photon fluxes in the two normal 
modes as functions of energy, rotational phase and the two angles $\chi$ 
and $\xi$ which characterize the viewing geometry. The Stokes parameter 
fluxes are also computed, taking into account the effects of both vacuum 
polarization and Stokes parameter rotation (see section \ref{subsubsec:stokesflux}). 
The polarization properties of the radiation, i.e. the linear polarization
fraction and the polarization angle are finally obtained for both phase-averaged 
and phase-resolved simulations.

\subsection{Second-order processes}
Throughout this paper, we considered magnetic Thomson scattering as the 
dominant source of opacity in the plasma. However, other second-order processes 
could be potentially relevant when strong magnetic fields are considered. 
In section \ref{subsection:radiativetransfer} we mentioned the role of double-Compton 
scattering in ensuring local thermal equilibrium at large optical depths 
deep in the fireball. Here we discuss the role of additional processes, 
such as thermal bremsstrahlung and photon splitting.

Since we considered non-relativistic particles ($kT\ll m_\echarge c^2$), 
electron-electron (positron-positron) bremsstrahlung turns out to be negligible. 
In fact, its contribution vanishes in the dipole approximation and it starts 
to be important only in the relativistic limit, at particle energies $\ga 
300$ keV \cite[the cross section is $\sim 2$ orders of magnitude smaller 
than the Thomson cross section at $\varepsilon\sim10$ keV, see][]{haug75}. 
Actually, in a pair plasma, where both electrons and positrons have comparable 
densities, the dominant contribution comes from electron-positron bremsstrahlung
\cite[][]{sven82,haug85a}. Unfortunately, no complete treatment of this process
in a strong magnetic field is currently available in the literature. Some 
considerations on its importance relative to scattering can nevertheless 
be made. In the free-field case, electron-positron emissivity is only slightly 
enhanced with respect to the electron-proton one \cite[by a factor $\sim2^{3/2}$;][]{sven82}. 
The cross section is $\sim0.5\,\sigmaT$, at $\varepsilon=1$ keV and $kT=10$ 
keV, and decreases for both increasing photon and particle energy \cite[see][]{haug85b}. 
Electron-proton bremsstrahlung has been investigated also in the strong-field 
limit by \citet[see also \citealt{laueretal83}]{lieu81}, who showed that 
the cross section for the extraordinary photons is strongly suppressed with 
respect to that for the O-mode, which, in turn, shows a behavior similar 
to that of the free-field limit. Assuming that \citet{sven82} result remains 
valid also at $B\ga B_\mathrm{Q}$, electron-positron bremsstrahlung cross 
section is likely to be smaller than the electron scattering one for both 
the ordinary and the extraordinary modes\footnote{The fact that the scattering 
depth is much greater than the free-free one for a plasma temperature $T\sim20$ 
keV was already noted by \citet{td95}, although no detailed discussion was 
provided.}. 

Photon splitting, on the other hand, can affect the photon spectra in a 
wide range of energies \cite[see e.g.][]{adleretal70,adler71}. The effects 
of this process, however, change radically according to the intensity of 
the magnetic field in which photons propagate, and the splitting of a photon 
in more than two photons is suppressed for $B\la B_\mathrm{Q}$ \cite[][]{bbbb70}. 
The expression for the photon splitting probability, that strongly depends 
on the polarization mode of the photons involved, has been discussed by 
\citet[see also \citealt{bulik98}]{ston79} in the weak-field limit and for 
low energies ($\varepsilon\ll m_\mathrm{e}c^2$). In particular, taking into 
account weak dispersive effects, the only allowed channel turns out to be 
that of an X-mode photon splitting in two O-mode photons\footnote{\citet{bulik98} 
reaches the same conclusion, specifying that the $\mathrm{O\rightarrow XX}$, 
$\mathrm{O\rightarrow OX}$ and $\mathrm{X\rightarrow OX}$ channels become 
important when the plasma contributions in the dielectric tensor dominate 
and in the high-energy range ($\varepsilon\ga m_\mathrm{e}c^2$) only.}. 
The maximum absorption coefficient\footnote{The photon splitting probability 
is maximized when the energy $\varepsilon'$ and $\varepsilon''$ of the two 
outgoing photons is half the energy $\varepsilon$ of the ingoing photon: 
$\varepsilon'=\varepsilon''=\varepsilon/2$.} of the process results
\begin{flalign} \label{equation:alphasplit}
\alpha_\mathrm{sp}(\mathrm{X}\rightarrow\mathrm{OO}) &= -\frac{\alpha_\mathrm{F}^3m_\mathrm{e}c}{60\pi^2\hbar}
\left(\frac{B\sin\theta_\mathrm{Bk}}{B_\mathrm{Q}}\right)^6\left(\frac{\varepsilon}{m_\mathrm{e}c^2}\right)^5 & \nonumber \\
 &\,\,\,\,\,\,\,\times \left[M_1(B/B_\mathrm{Q})\right]^2\,, &
\end{flalign}
where the complete expression of the function $M_1(B/B_\mathrm{Q})$ is given 
in \citet{ston79}. The dependencies on magnetic field and photon energy 
in equation (\ref{equation:alphasplit}) make clear that the effects of photon 
splitting are quite modest for photons with energies $1$--$100$ keV (the 
range of interest in this work) and for weak magnetic fields. On the other 
hand, although for super-critical fields one should consider also the contribution 
that comes from the splitting of a photon in more than two photons, the 
amplitude of the process goes as $\exp(-B/B_\mathrm{Q})$ \cite[][]{ston79}. 
A simple numerical estimate shows that the absorption coefficient associated 
to electron (positron) scattering $\alpha_\mathrm{sc}=n_\mathrm{e}\sigmaT$ 
typically exceeds the coefficient $\alpha_\mathrm{sp}$ by more than a factor 
$\sim3$ even at the highest photon energy we considered ($\varepsilon=100$ 
keV).

\subsection{Spectral analysis}
We presented the results of theoretical simulations for a template source 
endowed with a dipolar magnetic field, with polar intensity $B_\mathrm{p}=2\times10^{14}$ 
G. As it can be seen in Figure \ref{figure:spectidla}, the dominant contribution 
to the total spectrum comes from extraordinary photons, the ordinary photon 
flux being in general one order of magnitude (or more) smaller. This clearly 
follows from the fact that O-mode and X-mode photons have different scattering 
opacities, as discussed in section \ref{subsection:crosssections}. The cross
section for extraordinary photons is strongly supressed, by a factor $(\varepsilon/\varepsilonB)^2$,
with respect to that of ordinary ones. Nevertheless, as illustrated in Figure 
\ref{figure:fluxspectfortran}, O-mode photon contribution appears to increase 
at high energies ($\varepsilon\ga 50$ keV), as well as for lower magnetic 
field intensities, while the spectrum of X-mode ones is practically unchanged 
by varying the magnetic field strength. This suggests, in particular, that 
more energetic ordinary photons escape the fireball preferably far from 
the star surface, where the magnetic field is weaker, contrary to extraordinary 
ones. We found that the total spectrum can be well reproduced in terms of 
the superposition of two thermal components, in agreement with the results 
obtained by \citet{isretal08} in their study of the intermediate flares 
emitted during the SGR 1900+14 burst forest. We note that, although we considered 
only one illustrative case, corresponding to a bolometric temperature $T_\mathrm{b}=10$ 
keV and an angular opening $\phi_\maxrm-\phi_\minrm=90^\circ$, the temperatures 
and emitting area ratio we obtained for the two fitting components are compatible 
with observations. However, our results do not appear to support the suggestion 
by \citet{isretal08} according to which the soft and hard blackbodies are 
associated to photons coming from the O-mode and X-mode photospheres, respectively, 
as one can easily verify comparing Figures \ref{figure:spectidla} and \ref{figure:spectidlb}. 
The presence of two components in the spectral fit comes rather from the
broad distrubution of X-mode photons, which, according to their energy, 
escape the fireball atmosphere at different depths, where the temperature 
attains different values. This causes a flattening of the number 
flux at lower energies ($\varepsilon\la kT_\mathrm{b}$), as already noted 
by \citet{lyub02} and clearly visible in Figure \ref{figure:fluxspectfortran}.

Our model can also reproduce the pulsations observed in intermediate/giant 
flare decay tails. Actually, this problem has been extensively addressed 
by \cite{vput16}, who investigated, in particular, the geometry of the magnetar 
flare beaming, driven by relativistic outflows of charged particles \cite[][]{td95,td01}. 
According to their model, it is indeed the presence of these outflows that 
allows to explain the rotational modulation of the observed light curves 
as expected from the time evolution of the trapped fireball. They noted 
that, contrary of what observations show, the light curve modulation would 
change dramatically as the fireball shrinks, if only a localized emission 
region on the torus (and no beaming) is considered. However, the study of 
the time evolution of magnetar flares and how this can influence the shape 
of the light curves is outside the aims of our work. For this reason, in 
our model we focused only on the properties of the radiation emitted from 
a steady trapped-fireball, neglecting all the possible contributions coming 
from the advection of baryons due to relativistic outflows. Under these conditions,
we found that our model is able to account for the pulsations observed in the
decay tails of intermediate/giant flares, independently on whether the emitting
region of the torus-shaped fireball is limited azimuthally (model b) or not (model
a). The shape of the light curve and the pulsed fraction depend clearly on the viewing 
angles $\chi$ and $\xi$, as well as on the angular opening $\phi_\maxrm-\phi_\minrm$
in the case of model b, as shown in Figure \ref{figure:pulseprofile} and in the
bottom row of Figure \ref{figure:lightflux}. Moreover, a certain degree of beaming
turns out to be present also in our model, as visible e.g. in the angular distributions
plotted in Figure \ref{figure:angdistr} (which show that ordinary photons are preferentially
emitted along the local magnetic field direction) and in the top row of Figure \ref{figure:lightflux}.

\subsection{Polarization properties}
\begin{figure*}
\begin{center}
\includegraphics[width=17.5cm]{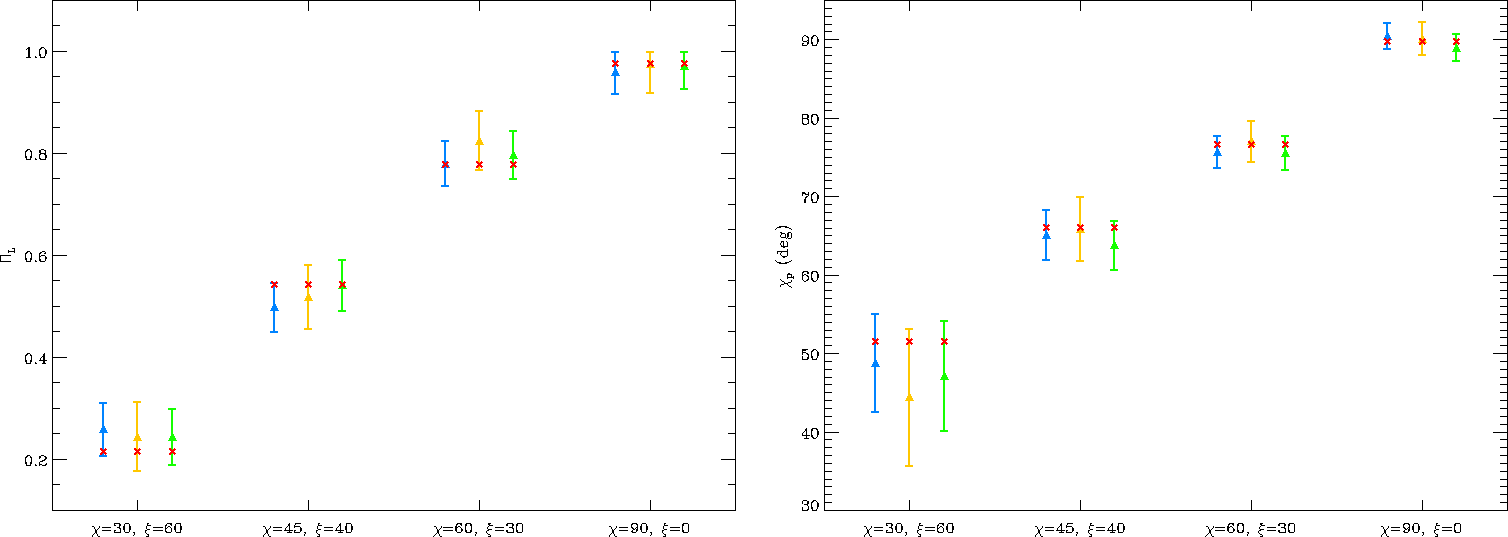}
\caption{Simulations of the phase-averaged response (triangles with error 
bars) of the XIPE (blue), IXPE (orange) and eXTP (green) polarimeters to 
the signal predicted by our theoretical model (in the case of model b) for 
an exposure time $t_\mathrm{exp}=1.737$ s, X-ray flux $F_\mathrm{X}=4.68\times10^{-7}$ 
erg cm$^{-2}$ s$^{-1}$ in the $1$--$10$ keV energy range and different viewing 
geometries (red crosses). The instrument specifications correspond to the 
respective baseline configurations.}
\label{figure:obssim}
\end{center}
\end{figure*}
Our work relies on the assumption that the plasma contributions 
to the dielectric tensor are negligible with respect to the vacuum terms. 
For this reason we did not consider vacuum resonance \cite[see e.g.][]{laiho03}, 
that occurs when plasma and vacuum contributions are comparable and influences 
the observed polarization signal switching the photon modes at an energy 
close to the resonant energy 
\cite[see][]{lyub02},
\begin{flalign}
\varepsilon_\mathrm{vr}&=\sqrt{6}m_\mathrm{e}c^2\bigg(\frac{2\pi T}{m_\mathrm{e}c^2}\bigg)^{1/4}\exp\bigg(-\frac{m_\mathrm{e}c^2}{2T}\bigg)\,, &
\end{flalign}
where $T$ is given by equation (\ref{equation:temperaturedistrib}). According 
to the original model by \citet{td95}, plasma effects in the fireball are 
essentially due to electron-positron pairs. The baryonic component inside 
the fireball is indeed much less important, since baryons are mostly advected 
away from the star surface through relativistic outflows \cite[see][]{vput16}. 
However, even at large optical depths, where the pair density is higher, 
the resonance energy results rather low ($\sim 1$ keV). This ensures that, 
over the entire $1$--$100$ keV energy range we considered, vacuum resonance 
can be safely neglected, assuming vacuum effects dominant over the plasma 
ones. Furthermore, the possible residual presence of baryons in the fireball 
is not even expected to modify photon polarization through scattering, since
the (Thomson) cross sections for photon scatterings onto baryons are suppressed 
with respect to those for scatterings onto pairs by a factor $\sim 10^{-6}$.
We note also that the possible effects of resonant scattering onto protons,
which occurs at the cyclotron energy $E_\mathrm{cp}=0.63(B/10^{14}\,\mathrm{G})$ keV, are not going
to affect the spectrum and the polarization properties in the energy range
we considered. In fact, for a polar magnetic field $B_\mathrm{p}=2\times10^{14}$
G, it is $E_\mathrm{cp}\la 1$ keV.

In the code both vacuum polarization effects and Stokes parameter 
rotation are accounted for (see section \ref{subsubsec:stokesflux}). We 
assumed that the adiabatic radius $r_\mathrm{a}$ is a sharp edge separating 
the adiabatic region (where the photon polarization vectors are locked to 
the star magnetic field direction) from the external region (where the polarization 
vector direction is frozen). Our results strongly indicate that magnetar 
flare radiation is highly polarized and dominated by extraordinary photons. 
In fact, as illustrated in the phase-resolved plots of Figure \ref{figure:PiLphres}, 
the linear polarization degree attains values higher than $80\%$ over almost 
all the entire $1$--$100$ keV energy range, dropping to about $70\%$ only 
at the highest energies. This decrease in $\Pi_\mathrm{L}$ is compatible 
with the increase of the ordinary photon contribution at higher energies 
we discussed in the previous section. Moreover, when an azimuthally limited 
emitting region is considered, radiation collected from the planar slices 
at the boundaries results in general even more polarized than the radiation 
with the same energy coming from the torus. 

As already pointed out in \citet{tavetal15}, phase-averaged simulations 
show more clearly the depolarizing effects of Stokes parameter rotation 
(see Figure \ref{figure:polobsPhAv}). This is essentially due to the fact 
that instruments give information about the Stokes parameters of each collected 
photon. Due to the Stokes parameter  rotation at the adiabatic radius, the 
average over the star rotational period generally reduce the observed polarization 
degree with respect to what one would see in a phase-resolved measurement, 
except for some favourable viewing geometries. In particular, the maximum 
$\Pi_\mathrm{L}$ (here nearly $100\%$) is attained for $\chi=90^\circ$, 
$\xi=0^\circ$, i.e. the case of an aligned rotator seen perpendicularly 
to the magnetic axis, where the effects of rotation are less important. 
Such a polarization signal is strong enough to be readily measurable by 
the new-generation, X-ray polarimeters currently under development. A plot 
of the simulated, phase-averaged response of XIPE, IXPE and eXTP to the 
signal predicted by our model, observed at different viewing geometries, 
is shown in Figure \ref{figure:obssim}. Here we refer to an event characterized 
by an X-ray flux $F_\mathrm{X}=4.68\times10^{-7}$ erg cm$^{-2}$ s$^{-1}$ 
in the $1$--$10$ keV energy range and an exposure time $t_\mathrm{exp}=1.737$ 
s, i.e. the values tabulated by \citet{isretal08} for the intermediate flare 
labelled IF1. Both polarization fraction and angle measurements recover 
the values expected from the theoretical model with an acceptable degree 
of accuracy (within 1 sigma). It can be noted that, while the errors on 
the polarization fraction are more or less the same for all the geometrical 
configurations considered, those on the polarization angle increase by decreasing 
the corresponding polarization degree. For $\Pi_\mathrm{L}\la 20\%$, polarization 
angle measurements turn out to be dominated by instrumental effects. In 
principle, polarization angle estimates could be useful to understand in 
which mode the collected radiation is polarized. However, as noted in previous 
works \cite[see e.g.][]{tavetal15,gonzetal16,mignetal17}, it should be taken 
into account that a polarization analysis alone does not suffice to this 
aim. In fact, the value of the polarization angle returned by the polarimeter 
depends on the orientation of its reference axis with respect to the projection 
of the star spin axis on the plane of the sky, that is a priori unknown. 
Nevertheless, as tested in the case of persistent emission \cite[see][]{tavetal14}, 
also for magnetar flares the oscillatory behavior of the polarization angle 
as a function of the rotational phase can be used to constrain the values 
of the viewing angles $\chi$ and $\xi$ (see Figure \ref{figure:chipolEnAv}).

\citet{yz15} have recently presented a model to investigate the polarization
properties of the radiation emitted during magnetar flare pulsating tails.
They calculated the radiative transfer in the fireball atmosphere using 
Monte Carlo simulations, in which they fixed the number of photons (5000) 
in two energy bands ($1$--$30$ keV and $30$--$100$ keV), assumed the star 
is an aligned rotator and collected photons for different inclination of 
the LOS with respect to the magnetic axis. QED effects were also considered 
in the ``sharp-edge'' approximation, as we did. While starting from the 
same initial conditions as in our work, their polarization fraction turns 
out to be quite modest, contrary to our findings. In fact, even if the collected 
radiation results mostly polarized in the extraordinary mode for all the 
considered viewing geometries, they obtained $\Pi_\mathrm{L}=30\%$ in the 
softer band (and only $10\%$ in the harder one) when the LOS is perpendicular 
to the magnetic axis, namely the configuration which should produce the 
largest phase-averaged polarization degree\footnote{For smaller angles between 
the LOS and the magnetic axis we obtain as well small polarization degrees 
due to the effects of Stokes parameter rotation.}. \citet{vput16} used as 
well a Monte Carlo code, fixing instead the computational time dedicated 
to each run rahter than the photon number. Although a complete analysis 
in this sense is outside their scopes, they also explored the polarization 
spectrum of the magnetar flare emission, finding that the ratio between 
ordinary and extraordinary photon intensities strongly depends on the outflow 
velocity. However, radiation appears to be largely dominated by ordinary 
photons, the two intensities becoming comparable only at high particle velocities 
($\ga0.8\,c$) and only for certain inclinations between the LOS and the 
magnetic axis. This could be explained by the fact that O-mode photons are 
tightly-coupled to the plasma deep in the fireball due to their large scattering 
opacity, so that they can be more easily advected by the relativistic outflow 
than X-mode one. Basically, in the case of sources for which an independent 
method to constrain the direction of the source rotation axis is possible, 
a polarization angle measurement can definitively disambiguate what polarization 
mode dominates the collected radiation, allowing to evaluate the actual 
role of relativistic outflows in characterizing magnetar flare emission. 

\section*{Acknowledgments}
We thank Luciano Nobili for his contribution during the early stages of
this investigation.

\appendix{} \label{sec:appendix}

\section{Local normal to the fireball surface} 
\label{appendix:localnormal}
Given a point $P$ on the fireball surface, characterized by the position 
vector
\begin{flalign} \label{equation:nthetaphi}
\boldsymbol{m}(\theta,\phi)&=r\left(\begin{array}{c}
\sin\theta\cos\phi \\ \sin\theta\sin\phi \\ \cos\theta
\end{array}\right)=R_\mathrm{max}\left(\begin{array}{c}
\sin^3\theta\cos\phi \\ \sin^3\theta\sin\phi \\ \sin^2\theta\cos\theta
\end{array}\right)\,, &
\end{flalign}
where $\theta$ and $\phi$ are related to $\Theta$ and $\Phi$ by equations
(\ref{equation:thetaphiGRpi}) and equation (\ref{equation:fieldlines}) has
been used, the surface normal $\boldsymbol{z}$ can be derived as
\begin{flalign} \label{equation:surfacenormalgen}
\boldsymbol{z}&=\frac{\boldsymbol{m}_\theta\times\boldsymbol{m}_\phi}{|\boldsymbol{m}_\theta\times\boldsymbol{m}_\phi|}\,, &
\end{flalign}
with 
\begin{flalign} \label{equation:nthetanphi}
\boldsymbol{m}_\theta&=\frac{\partial\boldsymbol{m}}{\partial\theta},\,\,\,\,\,\,\boldsymbol{m}_\phi=\frac{\partial\boldsymbol{m}}{\partial\phi}\,. &
\end{flalign}
Starting from equation (\ref{equation:nthetaphi}) and after some algebra
one obtains the components of $\boldsymbol{z}$ in the $\bdip$ frame,
\begin{flalign} \label{equation:surfacenormalpart}
\boldsymbol{z}&=\frac{1}{\sqrt{1+3\cos^2\theta}}\left(\begin{array}{c}
(1-3\cos^2\theta)\cos\phi \\ (1-3\cos^2\theta)\sin\phi \\ 3\sin\theta\cos\theta
\end{array}\right)\,. &
\end{flalign}

\section{Domain of the fireball terminator} 
\label{appendix:terminator}
The complete solution of the inequality (\ref{equation:terminatorinequality})
is given by the intersection of the solutions of $A<1$ and $A>-1$, where 
\mbox{$A=3\sin\theta\cos\theta\cos\eta/[\sin\eta(3\cos^2\theta-1)]$}.
Solving for $\theta$ the two equations $A=\pm 1$, one finds four distinct 
roots
\begin{flalign} \label{equation:inequalityroots}
t_{++}&\equiv\frac{3\cos\eta+\sqrt{\cos^2\eta+8}}{2\sin\eta}\,;\,\,\,t_{+-}\equiv\frac{3\cos\eta-\sqrt{\cos^2\eta+8}}{2\sin\eta} & \nonumber \\
t_{-+}&\equiv-t_{+-}\,;\,\,\,\,\,\,\,\,\,\,\,\,\,\,\,\,\,\,\,\,\,\,\,\,\,\,\,\,\,\,\,\,\,\,\,\,\,\,\,\,\,\,\,\,\,t_{--}\equiv-t_{++}\,. & \nonumber \\
\end{flalign}
In particular, solving separately $A<1$ and $A>-1$ leads to
\begin{flalign} \label{equation:Amin1}
&0<\theta<\minrm_{-+}^{(1)}\,\cup\,\,\maxrm_{-+}^{(1)}<\theta<\minrm_{--} ^{(2)}\,\cup\, & \nonumber \\
&\maxrm_{--}^{(2)}<\theta<\pi &  
\end{flalign}
and
\begin{flalign} \label{equation:Amagg-1}
&0<\theta<\minrm_{++}^{(1)}\,\cup\,\,\maxrm_{++}^{(1)}<\theta<\minrm_{+-}^{(2)}\,\cup\, & \nonumber \\
&\maxrm_{+-}^{(2)}<\theta<\pi\,, &  
\end{flalign}
respectively, where we defined
\begin{flalign} \label{equation:minmaxdef}
&\begin{array}{l}
\minrm_{\pm\pm}^{(1)}\,\equiv \min\left[\arctan(t_{\pm\pm}),\dfrac{1}{2}\arccos\left(-\dfrac{1}{3}\right)\right] \\
\ \\
\maxrm_{\pm\pm}^{(1)}\equiv \max\left[\arctan(t_{\pm\pm}),\dfrac{1}{2}\arccos\left(-\dfrac{1}{3}\right)\right] \\
\ \\
\minrm_{\pm\pm}^{(2)}\,\equiv \min\left[\pi+\arctan(t_{\pm\pm}),\pi-\dfrac{1}{2}\arccos\left(-\dfrac{1}{3}\right)\right] \\
\ \\
\maxrm_{\pm\pm}^{(2)}\equiv \max\left[\pi+\arctan(t_{\pm\pm}),\pi-\dfrac{1}{2}\arccos\left(-\dfrac{1}{3}\right)\right]\,.
\end{array}
&
\end{flalign}
In order to compute the intersection of the two solutions (\ref{equation:Amin1})
and (\ref{equation:Amagg-1}), it is necessary to sort the quantities given
in (\ref{equation:minmaxdef}) in increasing order,
\begin{flalign} \label{equation:sort}
(a_1,a_2,a_3,a_4)&\equiv \tt{sort}\normalfont(\minrm_{-+}^{(1)},\minrm_{++}^{(1)},\maxrm_{-+}^{(1)},\maxrm_{++}^{(1)}) & \nonumber \\
(b_1,b_2,b_3,b_4)&\equiv \tt{sort}\normalfont(\minrm_{--}^{(2)},\minrm_{+-}^{(2)},\maxrm_{--}^{(2)},\maxrm_{+-}^{(2)})\,. & 
\end{flalign}
In this way, one can write the ranges of $\theta$ for which the terminator
exists,
\begin{flalign} \label{equation:terminatordomain}
&\theta_\minrm<\theta<a_1\,\cup\,a_2<\theta<a_3\,\cup\,a_4<\theta\leq\pi/2\,\cup & \nonumber \\
&\pi/2<\theta<b_1\,\,\,\cup\,b_2<\theta<b_3\,\,\cup\,b_4<\theta<\theta_\maxrm\,. &
\end{flalign}
Actually, it can be shown that the two intervals $a_2<\theta<a_3$ and $b_2<\theta<b_3$
are present if and only if $a_2=\maxrm_{\pm+}^{(1)}$ and $b_2=\maxrm_{\pm-}^{(2)}$. 

\section{Coordinate transformation between the $\bdip$ and the LOS reference frames} 
\label{appendix:changeofbasis}
Given a vector $\boldsymbol{v}$ with components $(v_p,v_q,v_t)$ in the $\bdip$
reference frame, its components $(v_X,v_Y,v_Z)$ in the LOS frame are given
by the following change-of-basis transformation:
\begin{flalign} \label{equation:changeofbasis}
v_X&=v_pp_X+v_qq_X+v_tt_X & \nonumber \\
v_Y&=v_pp_Y+v_qq_Y+v_tt_Y & \nonumber \\
v_Z&=v_pp_Z+v_qq_Z+v_tt_Z\,, &
\end{flalign}
where the components of the unit vectors $\boldsymbol{p}$, $\boldsymbol{q}$
and $\boldsymbol{t}$ in the LOS frame are given in equations (\ref{equation:pqt}).

\label{lastpage}

\end{document}